\documentclass[hidelinks,journal]{IEEEtran}
\usepackage{amsmath,amssymb,amsfonts}
\usepackage{array}
\usepackage[caption=false,font=normalsize,labelfont=sf,textfont=sf]{subfig}
\usepackage{textcomp}
\usepackage{stfloats}
\usepackage{url}
\usepackage{verbatim}
\usepackage{graphicx}
\usepackage{cite}
\usepackage{xcolor}
\usepackage{orcidlink}
\usepackage[acronym,shortcuts]{glossaries}
\usepackage{bm}
\usepackage{relsize}
\usepackage{soul}
\usepackage{algorithm}
\usepackage{algcompatible}
\usepackage{mathtools}
\usepackage{bm}
\usepackage{enumitem}
\def\BibTeX{{\rm B\kern-.05em{\sc i\kern-.025em b}\kern-.08em
T\kern-.1667em\lower.7ex\hbox{E}\kern-.125emX}}

\newcommand{\trans}[0]{^{\mathsf{T}}}

\newcommand{\herm}[0]{^{\mathsf{H}}}

\newcommand{\Real}[1]{\Re\{{#1}\}}
\newcommand{\Imag}[1]{\Im\{{#1}\}}

\newacronym{CSI}{CSI}{channel state information}
\newacronym{LoS}{LoS}{line-of-sight}
\newacronym{NLoS}{NLoS}{non-LoS}
\newacronym{RPE}{RPE}{radar parameter estimation}
\newacronym{OTFS}{OTFS}{orthogonal time frequency space}
\newacronym{AFDM}{AFDM}{affine frequency division multiplexing}
\newacronym{CRLB}{CRLB}{Cram{\`e}r-Rao lower bound}
\newacronym{BCRLB}{BCRLB}{Bayesian Cram{\`e}r-Rao lower bound}
\newacronym{BBI}{BBI}{Bayesian bilinear inference}
\newacronym{AoA}{AoA}{angle-of-arrival}
\newacronym{SNR}{SNR}{signal-to-noise ratio}
\newacronym{ML}{ML}{maximum likelihood}
\newacronym{MIMO}{MIMO}{multiple-input multiple-output}
\newacronym{SISO}{SISO}{single-input single-output}
\newacronym{MUSIC}{MUSIC}{multiple signal classification}
\newacronym{MU}{MU}{multi-user}
\newacronym{ROOT-MUSIC}{ROOT-MUSIC}{ROOT multiple signal classification}
\newacronym{JCAS}{JCAS}{joint communication and sensing}
\newacronym{JCR}{JCR}{joint communications and radar}
\newacronym{ISAC}{ISAC}{integrated sensing and communications}
\newacronym{3D}{3D}{three-dimensional}
\newacronym{2D}{2D}{two-dimensional}
\newacronym{1D}{1D}{one-dimensional}
\newacronym{RX}{RX}{receive}
\newacronym{TX}{TX}{transmit}
\newacronym{BF}{BF}{beamformer}
\newacronym{ROI}{ROI}{region of interest}
\newacronym{mmWave}{mmWave}{millimeter-wave}
\newacronym{MF}{MF}{matched-filter}
\newacronym{DD}{DD}{delay-Doppler}
\newacronym{SotA}{SotA}{state-of-the-art}
\newacronym{ULA}{ULA}{uniform linear array}
\newacronym{QAM}{QAM}{quadrature amplitude modulation}
\newacronym{ISFFT}{ISFFT}{inverse symplectic finite Fourier transform}
\newacronym{SFFT}{SFFT}{symplectic finite Fourier transform}
\newacronym{ISI}{ISI}{inter-symbol interference}
\newacronym{AWGN}{AWGN}{additive white Gaussian noise}
\newacronym{MSE}{MSE}{mean-squared-error}
\newacronym{LMMSE}{LMMSE}{linear minimum mean square error}
\newacronym{RMSE}{RMSE}{root mean square error}
\newacronym{ESPRIT}{ESPRIT}{estimation of signal parameters via rotational invariant techniques}
\newacronym{OFDM}{OFDM}{orthogonal frequency division multiplexing}
\newacronym{OCDM}{OCDM}{orthogonal chirp division multiplexing}
\newacronym{BS}{BS}{base station}
\newacronym{UE}{UE}{user equipment}
\newacronym{JCDE}{JCDE}{joint channel and data estimation}
\newacronym{PDA}{PDA}{probabilistic data association}
\newacronym{PMF}{PMF}{probability mass function}
\newacronym{PBiGaBP}{PBiGaBP}{parametric bilinear Gaussian belief propagation}
\newacronym{PBiGAMP}{PBiGAMP}{parametric bilinear generalized approximate message passing}
\newacronym{GaBP}{GaBP}{Gaussian belief propagation}
\newacronym{GAMP}{GAMP}{generalized approximate message passing}
\newacronym{EP}{EP}{expectation propagation}
\newacronym{FT}{FT}{frequency-time}
\newacronym{DFT}{DFT}{discrete Fourier transform}
\newacronym{IDFT}{IDFT}{inverse discrete Fourier transform}
\newacronym{TD}{TD}{time domain}
\newacronym{wlg}{wlg}{without loss of generality}
\newacronym{CP}{CP}{cyclic prefix}
\newacronym{DAF}{DAF}{discrete affine Fourier}
\newacronym{DAFT}{DAFT}{discrete affine Fourier transform}
\newacronym{IDAFT}{IDAFT}{inverse discrete affine Fourier transform}
\newacronym{CPP}{CPP}{\textit{chirp-periodic} prefix}
\newacronym{IDZT}{IDZT}{inverse discrete Zak transform}
\newacronym{DZT}{DZT}{discrete Zak transform}
\newacronym{P/S}{P/S}{parallel-to-serial}
\newacronym{S/P}{S/P}{serial-to-parallel}
\newacronym{SBL}{SBL}{sparse Bayesian learning}
\newacronym{MPA}{MPA}{message passing algorithms}
\newacronym{EM}{EM}{expectation maximization}
\newacronym{sIC}{soft IC}{soft interference cancellation}
\newacronym{soft RG}{soft RG}{soft replica generation}
\newacronym{BG}{BG}{belief generation}
\newacronym{SGA}{SGA}{scalar Gaussian approximation}
\newacronym{CLT}{CLT}{central limit theorem}
\newacronym{PDF}{PDF}{probability density function}
\newacronym{QPSK}{QPSK}{quadrature phase-shift keying}
\newacronym{ICI}{ICI}{inter-carrier interference}
\newacronym{BER}{BER}{bit error rate}
\newacronym{DoF}{DoF}{degrees-of-freedom}
\newacronym{VGA}{VGA}{vector Gaussian approximation}
\newacronym{FD}{FD}{full-duplex}
\newacronym{SIC}{SIC}{self-interference cancellation}
\newacronym{NMSE}{NMSE}{normalized mean square error}
\newacronym{KL}{KL}{Kullback-Leibler}
\newacronym{SIMO}{SIMO}{single-input multiple-output}
\newacronym{MISO}{MISO}{multiple-input single-output}
\newacronym{i.i.d.}{i.i.d.}{independent and identically distributed}

\definecolor{carolinablue}{rgb}{0.6, 0.73, 0.89}

\begin{document}

\title{
{\large{\color{red}Please find the official IEEE published version of this article on IEEE Xplore {\color{blue}\href{https://ieeexplore.ieee.org/abstract/document/10794592}{[here]}} and cite as:}

{\color{cyan}K. R. R. Ranasinghe \textit{et al.}, "Joint Channel, Data, and Radar Parameter Estimation for AFDM Systems in Doubly-Dispersive Channels," in IEEE Transactions on Wireless Communications, vol. 24, no. 2, pp. 1602-1619, Feb. 2025, doi: 10.1109/TWC.2024.3510935.}

}

Joint Channel, Data and Radar Parameter Estimation for AFDM Systems in Doubly-Dispersive Channels}







\author{Kuranage Roche Rayan Ranasinghe\textsuperscript{\orcidlink{0000-0002-6834-8877}},~\IEEEmembership{Graduate Student Member,~IEEE,}\\
 Hyeon Seok Rou\textsuperscript{\orcidlink{0000-0003-3483-7629}},~\IEEEmembership{Graduate Student Member,~IEEE,} Giuseppe Thadeu Freitas de Abreu\textsuperscript{\orcidlink{0000-0002-5018-8174}},~\IEEEmembership{Senior Member,~IEEE,}\\ Takumi Takahashi\textsuperscript{\orcidlink{0000-0002-5141-6247}},~\IEEEmembership{Member,~IEEE,} and Kenta Ito\textsuperscript{\orcidlink{0009-0003-4625-281X}},~\IEEEmembership{Graduate Student Member,~IEEE \vspace{-5ex}}
\thanks{Kuranage Roche Rayan Ranasinghe, Hyeon Seok Rou and Giuseppe Thadeu Freitas de Abreu are with the School of Computer Science and Engineering, Constructor University (previously Jacobs University Bremen), Campus Ring 1, 28759 Bremen, Germany (emails: [kranasinghe, hrou, gabreu]@constructor.university). 

\indent Takumi Takahashi and Kenta Ito are with the Graduate School of Engineering, Osaka University, Suita 565-0871, Japan (emails: [takahashi@, k-ito@wcs.]comm.eng.osaka-u.ac.jp).

Part of this work was presented at the 2024 IEEE 21st Consumer Communications \& Networking Conference (CCNC) \cite{Furuta_CCNC_2024}.}
\indent }

\markboth{TO BE PUBLISHED IN IEEE TRANSACTIONS ON WIRELESS COMMUNICATIONS (ACCEPTED)}%
{Shell \MakeLowercase{\textit{et al.}}: A Sample Article Using IEEEtran.cls for IEEE Journals}


\maketitle

\begin{abstract}
We propose new schemes for \ac{JCDE} and \ac{RPE} in doubly-dispersive channels, such that \ac{ISAC} is enabled by \ac{UE} independently performing \ac{JCDE}, and \acp{BS} performing \ac{RPE}.
%
%
The contributed \ac{JCDE} and \ac{RPE} schemes are designed for waveforms known to perform well in doubly-dispersive channels, under a unified model that captures the features of either legacy \ac{OFDM}, \ac{SotA} \ac{OTFS}, and next-generation \ac{AFDM} systems.
The proposed \ac{JCDE} algorithm is based on a Bayesian \ac{PBiGaBP} framework first proposed for \ac{OTFS} and here shown to apply to all aforementioned waveforms, while the \ac{RPE} scheme is based on a new \ac{PDA} approach incorporating a Bernoulli-Gaussian denoising, optimized via \ac{EM}.
Simulation results demonstrate that \ac{JCDE} in \ac{AFDM} systems utilizing a single pilot per block significantly outperforms the \ac{SotA} alternative even if the latter is granted a substantial power advantage.
Similarly, the \ac{AFDM}-based \ac{RPE} scheme is found to outperform the \ac{OTFS}-based approach, as well as the \ac{SBL} technique, regardless of the waveform used.
\end{abstract}


\glsresetall

\vspace{-2ex}
\section{Introduction}
\label{sec:Introduction}


%
%
\IEEEPARstart{T}{he} building blocks of communications and radar systems have been traditionally designed separately and assumed to utilize orthogonal resources. 
\Ac{ISAC}\footnote{Also commonly referred to as \ac{JCAS} or \ac{JCR}.} \cite{Wild_Access_2021, Liu_JSAC_2022, Wang_CST_2023,GonzalezProcIEEE2024}, a recently-emerged technology seen by many as a key driver of future mobile communication systems, challenges that established concept by considering the simultaneous provision of radar and communications functionalities under the same architecture, having the advantages of more efficient use of resources, lower power consumption, decreased hardware costs and the potential for wide market penetration \cite{Liu_TC_2020, WymeerschPIMRC2021, WeiTIoT2023}.
While different perspectives on \ac{ISAC} exists, each with a particular paradigm -- $e.g.$ co-existence of communications and radar systems \cite{ZhengSPM2019} or dual-function approaches \cite{Tagliaferri_TWC_2023} -- the focus of this manuscript is on enabling sensing functionalities under wireless communications systems, also referred to as communication-centric \ac{ISAC}. 

%
%
Staying true to the integrative aspect of the proposition, earlier contributions to \ac{ISAC} focused on schemes employing existing communications waveforms.
In particular, many methods found in current literature seek to take advantage of certain features of channel estimation -- especially in higher frequency bands where less scattering, higher resolution and domain-specific sparsity enable the estimation of individual channel paths -- to incorporate radar-like functionalities, such as the estimation of radar parameters from echoes of the transmit signal \cite{GaudioTWC2020,Mohammed_BITS_2022, Geng_arxiv_2023, Gupta_OJCS_2024, Ranasinghe_ICASSP_2024}.

%
%
Examples of the latter are techniques based on the \ac{OFDM} waveform \cite{Geng_arxiv_2023, Gupta_OJCS_2024} which, however, indicated that the high \ac{ICI} can be a bottleneck that reduces robustness to high Doppler shifts present in doubly-dispersive environments, leading to severe degradation of performance.
This factor alone, introduced by the channel diagonals spreading into a decaying band \cite{Rou_SPM_2024}, severely hinders the use of \ac{OFDM} in high mobility scenarios \cite{Gaudio_TWC_2022}, as envisioned for 6G systems. 

%
%
Another potential candidate for the \ac{ISAC} paradigm is the \ac{OTFS} waveform, which was originally proposed in \cite{Hadani_WCNC_2017} as a novel \ac{2D} modulation scheme that embeds information directly on the \ac{DD} domain. 
Since the \ac{DD} domain has a direct parallel to common radar processing framework, \ac{OTFS} was quite popular at its inception, attracting much work that demonstrated its effectiveness for \ac{ISAC} \cite{GaudioTWC2020, Mohammed_BITS_2022, Geng_arxiv_2023, Ranasinghe_ICASSP_2024, Gupta_OJCS_2024}.
However, besides the significantly higher implementation complexity compared to \ac{OFDM}, which is inherent to its \ac{2D} modulation approach, another limitation of \ac{OTFS} is that it does not achieve the optimal diversity order in doubly-dispersive channels \cite{SurabhiTWC2019}.

%
%
In response to the above, \ac{AFDM} has recently emerged as a strong candidate for \ac{ISAC} \cite{BemaniAFDM_ICC_2021, Ni_ISWCS_2022, Bemani_TWC_2023, Bemani_WCL_2024}, gaining much attention thanks to its robust communications performance in high mobility scenarios and its ability to achieve optimal diversity order in doubly-dispersive channels.
The key idea of \ac{AFDM} is to employ the \ac{DAFT} to generate multiple orthogonal chirps that are then used to modulate the transmit signal.
Since \ac{AFDM} is otherwise similar to \ac{OFDM} in structure, it does not suffer from the high modulation complexity of \ac{OTFS}, and since the chirp parameters of \ac{AFDM} can be optimized to match the maximum Doppler shift of the doubly-dispersive channel, a full \ac{DD} representation of the channel can be achieved, giving the approach remarkable robustness to high mobility and/or frequency.

%
%
It has been in fact demonstrated that \ac{AFDM} has a superior \ac{BER} performance compared to \ac{OFDM}, achieving also \acp{BER} comparable to \ac{OTFS} while requiring lower overhead for channel estimation \cite{Bemani_TWC_2023} as a result of the \ac{1D} null-guard band required, as opposed to that of \ac{OTFS}, which is 2D.
In turn, evidence that \ac{AFDM} can be effectively utilized for \ac{ISAC} has also been given, for instance in \cite{Ni_ISWCS_2022} where a simple \ac{MF} approach was suggested.
Building on the latter, an improved \ac{AFDM}-based monostatic \ac{ISAC} system with inherent adaptation to self-interference mitigation was also designed in \cite{Bemani_WCL_2024}, albeit at the expense of a significant increase in computational complexity due to the exhaustive \ac{ML} grid search required for the \ac{RPE} procedure.

%
%
One aspect of the \ac{ISAC} concept that was often overlooked, and which recently has been addressed more frontally, however, is the impact of pilot overhead onto the overall approach, when designing the appropriate frame structure for optimum data decoding at a \ac{UE} being served.
It has been shown in \cite{Bemani_TWC_2023}, for instance, that a major advantage of \ac{AFDM} over \ac{OFDM} and \ac{OTFS} is the lighter requirement in the number of pilots, which in turn can be taken advantage of by enabling the \ac{BS} to perform \ac{RPE} without expensive \ac{SIC} techniques.
The trend to reduce pilot overhead suggests, therefore, the adoption of \ac{JCDE} technology, which has been shown to be very effective in allowing efficient channel estimation with very few pilots\footnote{The embedded channel estimation method proposed in \cite{Bemani_TWC_2023} requires a single pilot, but surrounded by a null-guard interval, such that the fraction of the \ac{AFDM} frame dedicated to payload is affected not just by the sole pilot but also the additional power that can be allocated to it.} and under strenuous conditions \cite{Ito_GLOBECOM_2020, Ito_ICC_2021, Iimori_TWC_2022, Ito_ICC_2023, TakahashiTWC_JCDE2024, Furuta_CCNC_2024,Akrout_ASILOMAR_2022}, significantly reduce the overhead of coherent wireless systems.
One example of the latter is the \ac{BBI} \ac{JCDE} scheme for \ac{OFDM} systems proposed in \cite{TakahashiTWC_JCDE2024}, which was shown to have the ability of handling channel aging and to entirely remove the need of pilots in favor of adequate coding at an equivalent rate.
Another is the work in \cite{Furuta_CCNC_2024}, where a Bayesian \ac{PBiGaBP} \ac{JCDE} scheme for \ac{OTFS} systems was proposed, which was shown to asymptotically approach the idealized scheme where perfect \ac{CSI} knowledge is available.

%
%
In view of all the above, we propose in this paper a combined and generalized \ac{JCDE}-enabled \ac{ISAC} solution for systems utilizing signals known to cope well with doubly-dispersive channels, including \ac{OFDM}, \ac{OTFS} and \ac{AFDM}. 
The proposed solution comprises two main contributions.
The first is a new \ac{JCDE} mechanism, designed with basis on the Bayesian \ac{PBiGaBP} originally proposed for \ac{OTFS} waveforms in \cite{Furuta_CCNC_2024}, which is shown to function also for the other modulation schemes under the general doubly-dispersive channel model developed in \cite{Rou_SPM_2024}. 
Comparisons of the \ac{SotA} \ac{OTFS} and the contributed \ac{AFDM} methods against a genie-aided linear \ac{GaBP} benchmark scheme, in which perfect \ac{CSI} is assumed, demonstrate that the proposed solution under \ac{AFDM} achieves the same performance of the \ac{OTFS} \ac{SotA}, both of which also closely approach the idealistic benchmark, indicating that there is no penalty for the reduced modulation complexity advantage obtained when utilizing \ac{AFDM} instead of \ac{OTFS}, and that the proposed scheme is effective in reducing pilot overhead.

The second contribution is a new \ac{RPE} scheme based on the vectorized \ac{PDA} approach \cite{Ito_ICC_2021, TakahashiTWC_JCDE2024} incorporating a Bernoulli-Gaussian denoiser tuned via \ac{EM} \cite{Vila_ASILOMAR_2011} which quickly converges to the appropriate distribution parameters.
Compared to the \ac{SotA} alternative adapted from \cite{Mehrotra_TCom_2023}, which is designed around the \ac{SBL} algorithm, the contributed method is shown to outperform the latter over any of the relevant waveforms, namely, \ac{OFDM}, \ac{OTFS} in \cite{Mehrotra_TCom_2023} and \ac{AFDM}. 
The technique also has an inherent and substantial computational cost advantage over \ac{ML}-based \ac{SotA} alternatives such as those in \cite{Bemani_TWC_2023} and \cite{Bemani_WCL_2024}.

The contributions of the article are summarized as follows:
\begin{itemize}
\item By concisely summarizing the unified \ac{OFDM}, \ac{OTFS} and \ac{AFDM} doubly-dispersive channel model from \cite{Rou_SPM_2024}, such that message-passing receivers for such waveforms can be designed under a common structure, we propose in Section \ref{sec:Generalized_JCDE_at_UE} a \ac{PBiGaBP}-based scheme for \ac{JCDE}, which applies to all the aforementioned waveforms.
\item We show that even if a single pilot per block is used, the proposed \ac{JCDE}-\ac{AFDM} system significantly outperforms both the \ac{OFDM} and \ac{OTFS} alternatives \cite{Furuta_CCNC_2024}, as well as the \ac{AFDM} method in \cite{Bemani_WCL_2024}, even when the latter is granted a substantial power advantage.
\item Finally, in Section \ref{sec:Proposed_PDA_MP}, a novel \ac{PDA}-based technique for \ac{RPE} in doubly-dispersive channels is proposed, which in addition to enabling monostatic implementation for \ac{AFDM} systems without requiring \ac{FD} technology, is shown to outperform the \ac{SotA} of \cite{Mehrotra_TCom_2023} for both \ac{OTFS} and \ac{AFDM} waveforms.
\end{itemize}

\vspace{-2ex}
\section{System Model}
\label{sec:SystemModel}

\begin{figure}[t]
  \centering
  \includegraphics[width=0.9\columnwidth]{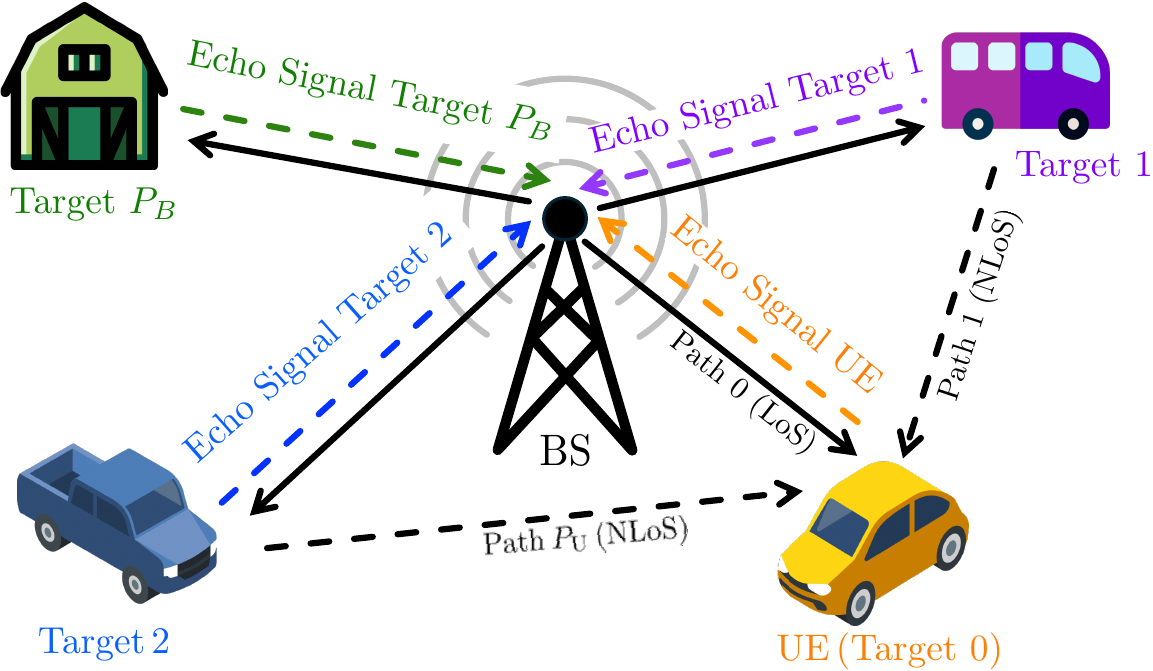}
  \vspace{-1ex}
  \caption[]{Illustration of an ISAC system with $P_\mathrm{B}$ target echoes received by a single antenna\footnotemark \ac{BS}, and $P_\mathrm{U}$ resolvable paths seen at the \ac{UE}.}
  \label{fig:ISAC_system_model}
  \vspace{-2ex}
  \end{figure}

We follow \cite{Bemani_WCL_2024} and consider an \ac{ISAC} scenario composed of a single-antenna\footnote{Both proposed algorithms can be extended straightforwardly to a \ac{SIMO} case, with some modifications due to spatial filtering/interference cancellation for which isotropic beamforming as done in \cite{Dehkordi_TWC23} could be used. Generalization to the \ac{MIMO} case is also possible, but requires the formulation and solution of a non-trivial \ac{TX} and \ac{RX} beamforming optimization problem. Such contribution falls outside the scope of this article and will be offered in a follow-up work.} \ac{BS} transmitting in downlink and taking advantage of first-order reflected (echo) signals to perform \ac{RPE} in order to detect the ranges and velocities of objects (targets) in its surroundings, while the \ac{UE} performs \ac{JCDE} as illustrated in Figure \ref{fig:ISAC_system_model}.
As depicted in the figure, it is assumed that a total of $P_\mathrm{B}$ target-originated echo signals reach the \ac{BS}, which might be either from \ac{UE} or other objects in the ambient.
In turn, each \ac{UE} is assumed to receive the signal transmitted by the \ac{BS} scattered onto $P_\mathrm{U}+1$ paths, with $p_\mathrm{U}=0$ denoting the \ac{LoS}, \ac{wlg}, and the remaining $P_\mathrm{U}\leq P_\mathrm{B}$ paths being \ac{NLoS} components of the multi-path channel.

For the sake of simplicity and again \ac{wlg}, we shall hereafter attain to a single user, since the \ac{MU} case has no impact onto the \ac{JCDE} procedure performed independently at each user, and is transparent to the \ac{RPE} algorithm, as long as the radar receiver has sufficient \ac{DoF} to resolve all $P_\mathrm{B}$ echoed signals it receives.
Also for simplicity, although we differentiate $P_\mathrm{B}$ and $P_\mathrm{U}$ in Figure \ref{fig:ISAC_system_model}, in the equations that follow, the number of paths/targets will be denoted by $P$, with the distinction between the two cases left to context.

\begin{figure*}[t!]
\setcounter{equation}{4}
\normalsize
\begin{equation}
\label{eq:diagonal_CP_matrix_def}
\mathbf{\Phi}_p \triangleq \text{diag}\bigg( [ \overbrace{e^{-j2\pi \phi_\mathrm{CP}(\ell_p)}, e^{-j2\pi \phi_\mathrm{CP}(\ell_p - 1)}, \dots, e^{-j2\pi \phi_\mathrm{CP}(2)}, e^{-j2\pi \phi_\mathrm{CP}(1)}}^{\ell_p \; \text{terms}}, \overbrace{1, 1, \dots, 1, 1}^{N - \ell_p \; \text{ones}}] \bigg) \in \mathbb{C}^{N \times N}.
\vspace{-2ex}
\end{equation}
%
%
\setcounter{equation}{5}
\begin{equation}
\label{eq:diagonal_Doppler_matrix_def}
\boldsymbol{\Omega} \triangleq \text{diag}\bigg([1,e^{-j2\pi /N},\dots,e^{-j2\pi (N-2) /N}, e^{-j2\pi (N-1) /N}]\bigg) \in \mathbb{C}^{N \times N}.
\end{equation}
\setcounter{equation}{0}
\hrulefill
\vspace{-3ex}
\end{figure*}

\addtocounter{footnote}{-1}
  
\footnotetext{Although it is possible to achieve $360^\circ$ coverage even in \ac{mmWave} bands with antenna arrays or omnidirectional antennas \cite{HongTAP2017, DengTAP2020, JoTIE2021}, for which Figure \ref{fig:ISAC_system_model} would apply as is, this illustration is not meant to be restricted to any specific band nor should it be interpreted as exclusive to the omnidirectional scenario, but rather can also be envisioned for sectorized scenarios as well \cite{GaudioTWC2020}.}

Finally, for the sake of enabling direct comparison with legacy (\ac{OFDM}) and \ac{SotA} (\ac{OTFS}) approaches, it shall be assumed that the \ac{BS} is capable of performing perfect \ac{SIC}\footnote{The problem of \ac{SIC} mitigation for the purpose of \ac{RPE} in \ac{AFDM} has already been addressed in \cite{Bemani_WCL_2024}, and solutions for the related problems of interference cancellation in \ac{OCDM}, or more generally for in-band interference cancellation in single antenna systems have also been proposed $e.g.$ in \cite{BomfinWCNC2019} and \cite{KhaledianTMTT2018}, respectively, which support the feasibility of the assumption.} prior to \ac{RPE}, to the advantage of the \ac{OFDM} and \ac{OTFS} approaches, since it has been shown \cite{Bemani_WCL_2024} that \ac{SIC} is more challenging and has higher computational complexity under those waveforms than under \ac{AFDM}.

\vspace{-2ex}
\subsection{Generalized Doubly-Dispersive Channel Model}
\label{sec:Generalized_Doubly-Dispersive_Channel_Model}
\vspace{-1ex}
Consider a general doubly-dispersive wireless channel model \cite{Rou_SPM_2024} characterized by $1$ \ac{LoS} and $P$ \ac{NLoS} propagation paths, with each $p$-th path comprising an uncorrelated complex fading gain $h_p \in \mathbb{C}$, a path delay $\tau_p \in [0,\tau_\text{max}]$ and a Doppler shift $\nu_p \in [-\nu_\text{max},\nu_\text{max}]$, respectively.
The delays and Doppler shifts of the doubly-dispersive channel are assumed to be bounded by a maximum delay $\tau_\text{max}$[s] and a maximum Doppler shift $\nu_\text{max}$[Hz], respectively, such that the \ac{TD} channel impulse response can be described by \cite{Bliss_Govindasamy_2013}
\vspace{-1ex}
\begin{equation}
\label{eq:doubly_dispersive_time_delay_channel}
h(t,\tau) \triangleq \sum_{p=0}^P h_p \cdot e^{j2\pi \nu_p t} \cdot \delta(\tau - \tau_p),
\end{equation}
where $t$ and $\tau$ are the continuous time and delays, respectively.

Leveraging equation \eqref{eq:doubly_dispersive_time_delay_channel}, the input-output relationship of the channel in the \ac{TD} can be obtained via the linear convolution of the transmit signal $s(t)$ and channel impulse response as described in \cite[eq. (10.69)]{Bliss_Govindasamy_2013}, which added with noise yields
\vspace{-1ex}
\begin{eqnarray}
\label{eq:doubly_dispersive_input_output_TD}
r(t) = s(t) * h(t,\tau) + w(t)  &&
\\ 
&&\hspace{-27.3ex} \triangleq \int\limits_{-\infty}^{+\infty} s(t-\tau) \bigg( \sum_{p=0}^P h_p \cdot e^{j2\pi \nu_p t} \cdot \delta(\tau - \tau_p) \bigg) d \tau + w(t), \nonumber\\[-2ex]
\nonumber
\end{eqnarray}
where $r(t)$ and $w(t)$ are the received signal and \ac{AWGN}, respectively.

Defining $r[n]$ and $s[n]$, with $n \in \{ 0,\dots,N-1 \}$, to be the sampled sequences of $r(t)$ and $s(t)$, respectively, with samples taken at a sufficiently high sampling rate $f_\mathrm{S} \triangleq \frac{1}{T_\mathrm{S}}$[Hz], the following discrete equivalent of equation \eqref{eq:doubly_dispersive_input_output_TD} can be obtained
\begin{equation}
\label{eq:doubly_dispersive_discrete_TD}
r[n] = \sum_{\ell = 0}^\infty s[n-\ell] \bigg( \sum_{p=0}^P h_p \cdot e^{j2\pi \frac{\nu_p}{f_\mathrm{S}} n} \cdot \delta\Big(\ell - \frac{\tau_p}{T_\mathrm{S}}\Big) \bigg) + w[n],
\vspace{-1ex}
\end{equation}
where $\ell$ is the normalized discrete delay index, and $T_\mathrm{S}$ the delay resolution as well as the sampling interval, such that $\tau = \ell\cdot T_\mathrm{S}$.

Finally, defining the normalized digital Doppler shift and the normalized discrete delay of the $p$-th path respectively as $f_p \triangleq \frac{N\nu_p}{f_\mathrm{S}}$ and $\ell_p \triangleq \frac{\tau_p}{T_\mathrm{S}}$, where the sampling rate $f_\mathrm{S}$ is chosen sufficiently high (via oversampling \cite{Rou_SPM_2024} if necessary)\footnote{The extension to incorporate normalized fractional delays if the chosen $f_\mathrm{S}$ is insufficient to render the integer assumption of the normalized delays is trivial and can be done via the approach suggested in \cite{Wu_TWC_2023}.
Subsequently, since the only difference would be a modification to $\mathbf{\Pi}$, both proposed algorithms could still be utilized as they are with no changes.} to ensure that $\ell_p - \lfloor \frac{\tau_p}{T_\mathrm{S}} \rceil \approx 0$, the discrete convolution described in equation \eqref{eq:doubly_dispersive_discrete_TD} can be rewritten in terms of the circular convolution
\vspace{-0.5ex}
\begin{equation}
\label{eq:channel_matrix_TD_general}
\mathbf{r} = \bigg( \sum_{p=0}^P h_p \cdot \mathbf{\Phi}_p \cdot \boldsymbol{\Omega}^{f_p} \cdot \mathbf{\Pi}^{\ell_p} \bigg) \cdot \mathbf{s} + \mathbf{w} = \mathbf{H} \cdot \mathbf{s} + \mathbf{w},
\vspace{-1ex}
\end{equation}
where the inclusion and posterior removal of a \ac{CP} of length $N_\mathrm{CP}$ to the sequence $s[n]$ is already taken into account \cite{Rou_SPM_2024}.

In equation \eqref{eq:channel_matrix_TD_general}, $\mathbf{r}\triangleq [r[0],\cdots,r[N-1]] \in \mathbb{C}^{N \times 1}$ and $\mathbf{s}\triangleq [s[0],\cdots,s[N-1]] \in \mathbb{C}^{N \times 1}$ are the transmit and received signal sample vectors; $\mathbf{w} \in \mathbb{C}^{N \times 1}$ denotes \ac{AWGN}; $\mathbf{H} \in \mathbb{C}^{N \times N}$ is the effective channel matrix; $\mathbf{\Phi}_p \in \mathbb{C}^{N \times N}$ described in equation \eqref{eq:diagonal_CP_matrix_def} is a diagonal matrix which captures the effect of the \ac{CP} onto the $p$-th channel path, where $\phi_\mathrm{CP}(n)$ denotes a phase function on the sample index $n \in \{ 0,\cdots,N-1 \}$ \cite{Rou_SPM_2024} depending on the specific waveform used; $\boldsymbol{\Omega} \in \mathbb{C}^{N \times N}$, described in equation \eqref{eq:diagonal_Doppler_matrix_def}, is a diagonal matrix containing the $N$ complex roots of one and $\mathbf{\Pi}\in \{0,1\}^{N \times N}$ is the forward cyclic shift matrix\footnote{To clarify, the matrix $\mathbf{\Pi}$ is such that $\bm{A}\cdot\mathbf{\Pi}^{\ell_p}$, with $\ell_p\in\mathbb{N}_0$, is a cyclic left-shifted version of the matrix $\bm{A}$, such that the first $\ell_p$ columns of $\bm{A}$ are moved to the positions of the last $\ell_p$ columns. Notice also that $\mathbf{\Pi}^0$ is the $N \times N$ identity matrix, such that $\bm{A}\cdot\mathbf{\Pi}^0=\bm{A}$.}, with elements given by
%
%
\vspace{-0.5ex}
\setcounter{equation}{6}
\begin{equation}
\label{eq:PiMatrix}
\pi_{i,j} = \delta_{i,j+1} + \delta_{i,j-(N-1)}\;\; \text{where}\;\; \delta _{ij} \triangleq
\begin{cases}
0 & \text{if }i\neq j,\\
1 & \text{if }i=j.
\end{cases}
\vspace{-0.25ex}
\end{equation}
%

Notice that the channel $\mathbf{H}$ implicitly defined in \eqref{eq:channel_matrix_TD_general} is a general representation of a doubly-dispersive channel \cite{Bliss_Govindasamy_2013}, which shall be exploited in the sequel to formulate estimation and detection models over various waveforms.

To that end, we consider the \ac{OFDM}, \ac{OTFS} and \ac{AFDM} waveforms, which are known to perform well in doubly-dispersive channels. 
Since a thorough comparison of the different individual features of each of them has been done in \cite{Rou_SPM_2024}, we limit ourselves here to briefly revise the models introduced thereby with the objective of obtaining a general mathematical formulation for the design of receivers/estimators for these waveforms under a common framework.

\vspace{-1ex}
\subsection{OFDM Signal Model}
\label{sec:OFDM_System_Model}

Let $\mathcal{C}\in\mathbb{C}^{Q\times 1}$ denote an arbitrary complex constellation of cardinality $Q$ and average energy $E_\mathrm{S}$, associated with a given digital modulation scheme.
In \ac{OFDM}, an input information vector $\mathbf{x} \in \mathcal{C}^{N\times 1}$ containing $N$ symbols taken from $\mathcal{C}$ are modulated into the transmit signal
\begin{equation}
\label{eq:OFDM_modulation}
\mathbf{s}_\text{OFDM} = \mathbf{F}_N\herm \cdot \mathbf{x} \in \mathbb{C}^{N \times 1},
\vspace{-0.5ex}
\end{equation}
where $\mathbf{F}_N$ denotes the $N$-point normalized \ac{DFT} matrix.

After undergoing the circular convolution with the doubly-dispersive channel, the corresponding \ac{OFDM} receive signal can be written in a form similar to that of equation \eqref{eq:channel_matrix_TD_general}, namely
\vspace{-0.5ex}
\begin{equation}
\label{eq:TD_OFDM_input_output}
\mathbf{r}_\text{OFDM} \triangleq \mathbf{H} \cdot \mathbf{s}_\text{OFDM} + \mathbf{w} \in \mathbb{C}^{N \times 1},
\vspace{-0.5ex}
\end{equation}
which can be demodulated into
\vspace{-0.5ex}
\begin{equation}
\label{eq:OFDM_demodulation}
\mathbf{y}_\text{OFDM} = \mathbf{F}_N \cdot \mathbf{r}_\text{OFDM} \in \mathbb{C}^{N \times 1}.
\vspace{-0.5ex}
\end{equation}

From equations \eqref{eq:TD_OFDM_input_output} and \eqref{eq:OFDM_demodulation}, one obtains a straightforward input-output relationship for \ac{OFDM} modulation, namely
\vspace{-0.5ex}
\begin{equation}
\label{eq:OFDM_input_output}
\mathbf{y}_\text{OFDM} = \mathbf{G}_\text{OFDM} \cdot \mathbf{x} + \tilde{\mathbf{w}} \in \mathbb{C}^{N \times 1},
\vspace{-0.5ex}
\end{equation}
where $\tilde{\mathbf{w}} \triangleq \mathbf{F}_N \mathbf{w} \in \mathbb{C}^{N \times 1}$ is an equivalent \ac{AWGN} vector with the same statistics as $\mathbf{w}$, while $\mathbf{G}_\text{OFDM}$ is the effective \ac{OFDM} channel, defined by
\vspace{-0.5ex}
\begin{equation}
\label{eq:OFDM_effective_channel}
\mathbf{G}_\text{OFDM} \triangleq \sum_{p=0}^P h_p \cdot \mathbf{F}_N \cdot (\boldsymbol{\Omega}^{f_p} \cdot \mathbf{\Pi}^{\ell_p}) \cdot \mathbf{F}_N\herm,
\end{equation}
with the \ac{CP} phase matrices $\mathbf{\Phi}_p$ appearing in equation \eqref{eq:channel_matrix_TD_general} reduced to identity matrices \cite{Rou_SPM_2024}; $i.e.,$ $\phi_\mathrm{CP}(n) = 0$ in equation \eqref{eq:diagonal_CP_matrix_def} since there is no phase offset in the \ac{CP} of \ac{OFDM} signals.

\vspace{-2ex}
\subsection{OTFS Signal Model}
\label{sec:OTFS_System_Model}

Consider an \ac{OTFS} system in which an input information matrix $\mathbf{X}\in \mathcal{C}^{K\times M}$, containing $K\times M$ complex symbols taken from an arbitrary constellation $\mathcal{C}$, is modulated into the transmit signal\footnote{For simplicity, we assume that all pulse-shaping operations utilize rectangular waveforms, such that the corresponding sample matrices can be reduced to identity matrices.}
\vspace{-0.5ex}
\begin{equation}
\label{eq:TD_transmit_matrix_vectorized}
\mathbf{s}_\text{OTFS} \triangleq \text{vec}(\mathbf{S}) = (\mathbf{F}_M\herm \otimes \mathbf{I}_K) \cdot \mathbf{x} \in \mathbb{C}^{KM\times 1},
\end{equation}
where $\text{vec}(\cdot)$ denotes the vectorization of a matrix, obtained by stacking the columns; the symbol $\otimes$ denotes Kronecker product; and $\mathbf{S}$ is a \ac{TD} symbol matrix obtained\footnote{Equivalently, $\mathbf{S}$ can be obtained as the Heisenberg transform of the \ac{ISFFT} of $\mathbf{X}$, $i.e.$ $\mathbf{S} = \mathbf{F}_K\herm \mathbf{X}_\text{FT}$ with $\mathbf{X}_\text{FT} \triangleq \mathbf{F}_K \mathbf{X} \mathbf{F}_M\herm \in \mathbb{C}^{K\times M}$.} from the \ac{IDZT} of $\mathbf{X}$, $i.e.$ \cite{Hadani_WCNC_2017}
\begin{equation}
\label{eq:TD_transmit_matrix}
\mathbf{S} = \mathbf{X} \mathbf{F}_M\herm  \in \mathbb{C}^{K\times M}.
\end{equation}

We highlight that the notation in equation \eqref{eq:TD_transmit_matrix_vectorized} is in line with the strategy described $e.g.$ in \cite{Raviteja_TWC_2018}, whereby the \ac{OTFS} signals are first vectorized and then appended with a \ac{CP} of length $N_\mathrm{CP}$ in order to eliminate inter-frame interference, in similarity with \ac{OFDM}.
Taking advantage of such similarity, and in order to allow for direct comparisons between the two waveforms, we shall hereafter set $K\times M = N$.

After transmission over the doubly-dispersive channel $\mathbf{H}$, the \ac{OTFS} receive signal can be modeled as
\begin{equation}
\mathbf{r}_\text{OTFS} \triangleq \mathbf{H} \cdot \mathbf{s}_\text{OTFS} + \mathbf{w}    \in \mathbb{C}^{N\times 1},
\label{eq:OTFS_received vector_in_TD}
\end{equation}
which is analogous to the \ac{OFDM} model of equation \eqref{eq:OFDM_input_output}. 

Unlike \ac{OFDM}, however, the detection of the information symbols in $\mathbf{x}$ from  $\mathbf{r}_\text{OTFS}$ requires reversing the vectorization and the \ac{IDZT} operations employed in the construction of $\mathbf{s}_\text{OTFS}$, resulting in a distinct effective channel.

In particular, let $\mathbf{R} \triangleq \text{vec}^{-1}(\mathbf{r}_\text{OTFS}) \in \mathbb{C}^{K \times M}$, where $\text{vec}^{-1}(\cdot)$ denotes the de-vectorization operation whereby a vector of size $KM \times 1$ is reshaped into a matrix of size $K \times M$, and consider the corresponding \ac{DZT}\footnote{Equivalently, $\mathbf{Y}$ can be obtained as the \ac{SFFT} of the Wigner transform $\mathbf{Y}_\text{FT} \triangleq \mathbf{F}_K \mathbf{R}$ of the matrix $\mathbf{R}$, which yields $\mathbf{Y} = \mathbf{F}_K\herm \mathbf{Y}_\text{FT} \mathbf{F}_M\in \mathbb{C}^{K \times M}$.}
\begin{equation}
\label{eq:DD_rec_sig_after_SFFT}
\mathbf{Y}  =  \mathbf{R} \mathbf{F}_M \in \mathbb{C}^{K \times M}.
\end{equation}

The corresponding demodulated \ac{OTFS} signal is given by
\begin{equation}
\label{eq:DD_demodulation}
\mathbf{y}_\text{OTFS} \triangleq \text{vec}(\mathbf{Y}) = (\mathbf{F}_M \otimes \mathbf{I}_K) \cdot \mathbf{r}_\text{OTFS} \in \mathbb{C}^{N\times 1},
\end{equation}
which in turn can be rewritten in the form of the direct input-output relation
\begin{equation}
\label{eq:DD_input_output_relation}
\mathbf{y}_\text{OTFS} = \mathbf{G}_\text{OTFS} \cdot \mathbf{x} + \tilde{\mathbf{w}}_\text{OTFS}\in \mathbb{C}^{N \times 1},
\end{equation}
where $\tilde{\mathbf{w}}_\text{OTFS}\triangleq (\mathbf{F}_M \otimes \mathbf{I}_K) \cdot \mathbf{w} \in \mathbb{C}^{N \times 1}$ is an equivalent \ac{AWGN} vector with the same statistics as $\mathbf{w}$, while $\mathbf{G}_\text{OTFS} \in \mathbb{C}^{N \times N}$ is the effective \ac{OTFS} channel in the \ac{DD} domain, defined by
\begin{equation}
\label{eq:DD_domain_effective_channel}
\mathbf{G}_\text{OTFS} \triangleq \sum_{p=0}^P h_p \cdot (\mathbf{F}_M \otimes \mathbf{I}_K) \cdot (\boldsymbol{\Omega}^{f_p} \cdot \mathbf{\Pi}^{\ell_p}) \cdot (\mathbf{F}_M\herm \otimes \mathbf{I}_K),
\end{equation}
in which, similar to the \ac{OFDM} waveform, the \ac{CP} phase matrices $\mathbf{\Phi}_p$ have been reduced to identity matrices \cite{Rou_SPM_2024}.

Comparing equations \eqref{eq:OFDM_effective_channel} and \eqref{eq:DD_domain_effective_channel}, one can appreciate how the channel modeling approach of \cite{Rou_SPM_2024} elucidates both the similarity in form, and the distinction in effect between the \ac{OFDM} and \ac{OTFS} waveforms in doubly-dispersive channels.

\subsection{AFDM Signal Model}
\label{sec:AFDM_System_Model}

Finally, let us describe the input-output relationship of the \ac{AFDM} waveform associated with a  doubly-dispersive channel.
Similarly to the above, letting $\mathbf{x}$ denote the information vector with symbols taken from the constellation $\mathcal{C}$, the corresponding \ac{AFDM} modulated transmit signal is given by its \ac{IDAFT}, $i.e.$
\begin{equation}
\label{eq:AFDM_moduation}
\mathbf{s}_\text{AFDM} = (\mathbf{\Lambda}_1\herm \mathbf{F}_{N}\herm \mathbf{\Lambda}_2\herm) \cdot \mathbf{x} \in \mathbb{C}^{N \times 1},
\end{equation}
with the matrices $\mathbf{\Lambda}_i$ defined as
%
%
\begin{figure*}[t!]
  \setcounter{equation}{21}
  \normalsize
  \begin{equation}
  \label{eq:AFDM_diagonal_CP_matrix_def}
  \bm{\varPhi}_p \triangleq \text{diag}\bigg( [ \overbrace{e^{-j2\pi c_1 (N^2-2N\ell_p)}, e^{-j2\pi c_1 (N^2-2N(\ell_p-1))}, \dots, e^{-j2\pi c_1 (N^2-2N)}}^{\ell_p \; \text{terms}}, \overbrace{1, 1, \dots, 1, 1}^{N - \ell_p \; \text{ones}}] \bigg) \in \mathbb{C}^{N \times N}.
  \vspace{-1ex}
  \end{equation}
  \setcounter{equation}{20}
  \hrulefill
  \vspace{-3ex}
\end{figure*}
%
%
\begin{equation}
\label{eq:lambda_def}
\mathbf{\Lambda}_i \triangleq \text{diag}\big(\big[1, \cdots, e^{-j2\pi c_i n^2}, \cdots, e^{-j2\pi c_i (N-1)^2}\big]\big) \in \mathbb{C}^{N \times N},
\end{equation}
where the constants $c_1$ and $c_2$ are chosen appropriately to match the maximum Doppler of the channel, and include the insertion of a \ac{CPP} to mitigate the effects of multipath propagation \cite{Bemani_TWC_2023}. 

It was shown in \cite{Rou_SPM_2024} that after going through a doubly-dispersive channel, an \ac{AFDM} modulated symbol vector as given in equation \eqref{eq:AFDM_moduation} with the inclusion of a \ac{CPP} as described in \cite{Bemani_TWC_2023}, can be modeled similarly to equation \eqref{eq:channel_matrix_TD_general}, only with the \ac{CP} matrix $\bm{\Phi}_p$ of equation \eqref{eq:diagonal_CP_matrix_def} replaced by the corresponding \ac{CPP} matrix $\bm{\varPhi}_p$ as explicitly given in equation \eqref{eq:AFDM_diagonal_CP_matrix_def} ($i.e.,$ $\phi_\mathrm{CP}(n) = c_1 (N^2 - 2Nn)$ in equation \eqref{eq:diagonal_CP_matrix_def}), such that the corresponding received signal is given by
%
%
\setcounter{equation}{22}
\begin{equation}
\label{eq:AFDM_received vector_in_TD}
\mathbf{r}_\text{AFDM} \triangleq \mathbf{H} \cdot \mathbf{s}_\text{AFDM} + \mathbf{w} \in \mathbb{C}^{N\times 1}.
\end{equation}
%

Demodulating the signal in equation \eqref{eq:AFDM_received vector_in_TD} yields
\begin{equation}
\mathbf{y}_\text{AFDM} = (\mathbf{\Lambda}_2 \mathbf{F}_{N} \mathbf{\Lambda}_1) \cdot \mathbf{r}_\text{AFDM} \in \mathbb{C}^{N\times 1},
\label{eq:AFDM_demodulation}
\end{equation}
which in turn can be expressed as
\begin{equation}
\mathbf{y}_\text{AFDM} = \mathbf{G}_\text{AFDM} \cdot \mathbf{x} + \tilde{\mathbf{w}}_\text{AFDM} \in \mathbb{C}^{N\times 1},
\label{eq:DAF_input_output_relation}
\end{equation}
where $\tilde{\mathbf{w}}_\text{AFDM} \triangleq (\mathbf{\Lambda}_2 \mathbf{F}_{N} \mathbf{\Lambda}_1) \cdot \mathbf{w} \in \mathbb{C}^{N\times 1}$ is an equivalent \ac{AWGN} vector with the same statistics as $\mathbf{w}$, and $\mathbf{G}_\text{AFDM} \in \mathbb{C}^{N\times N}$ in the effective \ac{AFDM} channel matrix, given by
\begin{equation}
\label{eq:DAF_domain_effective_channel}
\mathbf{G}_\text{AFDM} \triangleq \sum_{p=0}^P h_p \cdot (\mathbf{\Lambda}_2 \mathbf{F}_{N} \mathbf{\Lambda}_1) \cdot (\bm{\varPhi}_p \cdot \boldsymbol{\Omega}^{f_p} \cdot \mathbf{\Pi}^{\ell_p}) \cdot (\mathbf{\Lambda}_1\herm \mathbf{F}_{N}\herm \mathbf{\Lambda}_2\herm),
\end{equation}
where we emphasize that $\bm{\varPhi}_p$ is as in equation \eqref{eq:AFDM_diagonal_CP_matrix_def}.

Once again, it is clear that equations \eqref{eq:OFDM_effective_channel}, \eqref{eq:DD_domain_effective_channel} and \eqref{eq:DAF_domain_effective_channel} have the same structure, as do the input-output relationships described by equations \eqref{eq:OFDM_input_output}, \eqref{eq:DD_input_output_relation} and \eqref{eq:DAF_input_output_relation}, such that the \ac{JCDE} and \ac{RPE} techniques proposed in the sequel are applicable to \ac{OFDM}, \ac{OTFS} and \ac{AFDM} waveforms alike.

\section{Joint Channel and Data Estimation}
\label{sec:Generalized_JCDE_at_UE}

In this section, we introduce the proposed \ac{JCDE} method for \ac{OFDM}, \ac{OTFS} and \ac{AFDM} systems, under the general system model described in the previous section, and compare their relative performances.
For the sake of convenience, we shall express the effective channels for the three waveforms given in equations \eqref{eq:OFDM_effective_channel}, \eqref{eq:DD_domain_effective_channel} and \eqref{eq:DAF_domain_effective_channel} in the form
\begin{equation}
\label{eq:final_generalized_input_output_relation}
\mathbf{y} = \sum_{p=0}^P h_p \cdot \mathbf{\Gamma}_p \cdot \mathbf{x} + \tilde{\mathbf{w}} \in \mathbb{C}^{N\times 1},
\end{equation}
where the matrices $\mathbf{\Gamma}_p$ captures the long-term delay-Doppler statistics of the channel, which are assumed to be known\footnote{According to \cite{MatzTWC2005,MishraTWC2022,YangTVT2024}, the delay and Doppler shifts can be assumed constant during multiple frame transmissions, enabling sporadic estimation of $\mathbf{\Gamma}_p, \forall p$ via \ac{RPE} schemes utilizing pilots such as the method described in the latter part of Section \ref{sec:Proposed_PDA_MP}. However, the addition of this procedure into the \ac{JCDE} scheme transforms this into an \ac{RPE}-aided \ac{JCDE} method ($i.e.,$ user-centric \ac{RPE}) which while consolidating and combining the seperate aspects of \ac{ISAC}, deserves further attention and will be addressed in a follow-up article.}.

From equation \eqref{eq:final_generalized_input_output_relation}, it is evident that each received signal in $\mathbf{y}$ can be described by
\begin{equation}
y_n = \sum_{p=0}^P \sum_{m=0}^{N-1} h_p \cdot \gamma_{p:n,m} \cdot x_m + \tilde{w}_n,
\label{eq:elementwise_final_input_output_relation}
\end{equation}
where we slightly deviate from the notation employed earlier in equation \eqref{eq:doubly_dispersive_discrete_TD} by denoting the $m$-th information symbol as $x_m$ instead of $x[m]$, and accordingly the $n$-th receive signal sample by $y_n$ instead of $y[n]$, for future convenience.

\subsection{JCDE\,via\,Parametric Bilinear Gaussian Belief Propagation}
\label{sec:SotA_GAMP_GaBP}

We have first proposed the \ac{PBiGaBP} framework for \ac{JCDE} under \ac{OTFS} signaling in \cite{Furuta_CCNC_2024}. 
Thanks to the generalized channel model described above, however, it is obvious that the same technique applies to \ac{OFDM} and \ac{AFDM} as well.
We therefore proceed to succinctly describe the approach, referring the reader to \cite{Furuta_CCNC_2024} for further details.

For starters, referring to Figure \ref{fig:ISAC_frame_design}, it is initially assumed\footnote{Later we shall adopt the strategy proposed in \cite{Bemani_TWC_2023} and consider, for the \ac{AFDM} case, the alternative of utilizing only one pilot followed by a null-guard interval of length $B-1$ among the $N$ symbols in $\mathbf{x}$, for the purpose of enabling the \ac{RPE} scheme of Section \ref{sec:Proposed_PDA_MP}.} that in any of the three systems compared, a block of $B$ pilots\footnote{We use the first column of the $B\times B$ Zadoff-Chu sequence \cite{Furuta_CCNC_2024} as our pilot column.} is embedded in the $N\times 1$ symbol vector
$\mathbf{x}$.

%
%

We shall also clarify that although the placement of pilots amidst payload (data) symbols can be optimized under various criteria \cite{HamiltonTWC2011}, in the context of our discussion where issues such as distortions due hardware imperfection \cite{HeTSP2011}, pilot contamination \cite{YangTWC2018}, and channel aging \cite{TakahashiTWC_JCDE2024} are not considered, the location of pilots has no effect on the performance of the \ac{JCDE}.
Suffice it therefore, to consider the amount of power allocated to pilots, relative to that allocated to data symbols.

With these issues addressed, the \ac{PBiGaBP} framework for \ac{JCDE} introduced here can be summarized as follows:
\begin{itemize}[leftmargin=2ex]
\setlength\itemsep{0.1ex}
\item \Ac{sIC}, which consists of removing the inter-symbol interference from $\mathbf{y}$, by using temporary estimates (\emph{soft replicas}) of $\mathbf{h}$ and $\mathbf{x}$ generated in the previous iteration;
\begin{figure}[H]
  \centering
  \includegraphics[width=0.9\columnwidth]{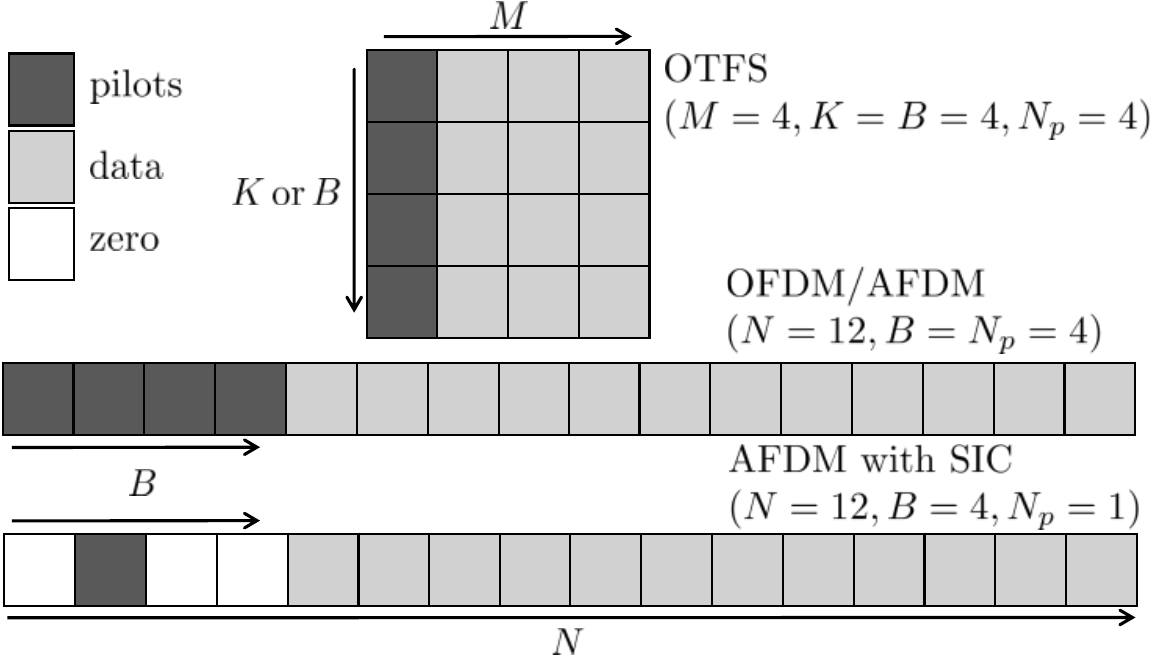}
  \vspace{-2ex}
  \caption{Illustration of pilot allocation schemes in OFDM, OTFS and AFDM.}
  \label{fig:ISAC_frame_design}
\end{figure}
\item \Ac{BG}, whereby the likelihood (\textit{beliefs}) of the estimates of $\mathbf{h}$ and $\mathbf{x}$ are obtained from $\mathbf{y}$ after \ac{sIC} and under \ac{SGA};
\item \Ac{soft RG}, whereby soft replicas are generated from the beliefs via conditional expectations.
\end{itemize}

The signal processing operations corresponding to each of these steps are described below.

\subsubsection{Soft IC}

Let us consider the $i$-th iteration of the algorithm, and denote the soft replicas of the $m$-th symbol and the $p$-th channel gain associated with the $n$-th receive signal $y_n$, at the previous iteration respectively by $\hat{x}_{n,m}^{(i-1)}$ and $\hat{h}_{n,p}^{(i-1)}$.
Then, the \acp{MSE} of these estimates computed for the $i$-th iteration are given by
\begin{subequations}
\begin{equation}
\hat{\sigma}^{2(i)}_{x:{n,m}} \triangleq \mathbb{E}_{x} \big[ | x - \hat{x}_{n,m}^{(i-1)} |^2 \big]= E_\mathrm{S} - |\hat{x}_{n,m}^{(i-1)}|^2, \forall (n,m),
\label{eq:MSE_x_m}
\end{equation}
\begin{equation}
\hat{\sigma}^{2(i)}_{h:{n,p}} \triangleq \mathbb{E}_{h} \big[ | h - \hat{h}_{n,p}^{(i-1)} |^2 \big] = \sigma_h^2 -|\hat{h}_{n,p}^{(i-1)} |^2, \forall (n,p),
\label{eq:MSE_h_p}
\end{equation}
where $\mathbb{E}_{x}$ refers to expectation over the all possible symbols $x$ in the constellation $\mathcal{C}$, while $\mathbb{E}_{h}$ refers to expectation over all possible outcomes of $h\sim\mathcal{CN}(0,\sigma^2_h)$, respectively\footnote{Although the average power of each channel path may be generally different, especially in sub-6GHz channels where the propagation lengths of different paths can vary from meters to kilometers, it is commonly assumed in high-frequency (\ac{mmWave} and Terahertz) channels, where ranges are much shorter, that the gains of all paths follow the same distribution \cite{Ni_ISWCS_2022,Bemani_TWC_2023,Bemani_WCL_2024}.}.
\end{subequations}

At a given $i$-th iteration, the objective of the \ac{sIC} step is to compute the symbol- and channel-centric replicas $\tilde{y}_{x:{m,n}}^{(i)}$ and $\tilde{y}_{h:{p,n}}^{(i)}$ and the corresponding variances $\tilde{\sigma}^{2(i)}_{x:{m,n}}$ and $\tilde{\sigma}^{2(i)}_{h:{p,n}}$, using the soft estimates $\hat{x}_{n,m}^{(i-1)}$ and $\hat{h}_{n,p}^{(i-1)}$ and their variances $\hat{\sigma}_{x:{n,m}}^{2(i-1)}$ and $\hat{\sigma}^{2(i-1)}_{h:{n,p}}$ obtained in the previous iteration.

In view of equation \eqref{eq:elementwise_final_input_output_relation}, we straightforwardly have
\begin{equation}
\label{eq:soft_IC_process}
\tilde{y}_{x:{m,n}}^{(i)} = y_n - \sum_{p=0}^P \sum_{q \neq m}^{N-1} \hat{h}_{n,p}^{(i-1)} \cdot \gamma_{p:n,q} \cdot \hat{x}_{n,q}^{(i-1)}.
\end{equation}

These \ac{sIC} receive signal replicas follow Gaussian \acp{PDF}, namely
\begin{equation}
\tilde{y}_{x:{m,n}}^{(i)} \sim p_{\text{y}_n | x_m}(\text{y}_n | x_m) \propto \text{exp}\bigg(\! - \frac{|\text{y}_n\! -\! \tilde{\gamma}_{x:{m,n}}^{(i)} x_m|^2}{\tilde{\sigma}^{2(i)}_{x:{m,n}}} \bigg),
\label{eq:conditional_PDF_y_nm_x}
\end{equation}
where $\text{y}_n$ is an auxiliary variable and $\tilde{\gamma}_{x:{m,n}}^{(i)}$ are the corresponding \ac{sIC} effective channel gains, defined as
\begin{equation}
\label{eq:effective_channel_gain_a_in_PDF}
\tilde{\gamma}_{x:{m,n}}^{(i)} \triangleq \sum_{p=0}^P \hat{h}_{n,p}^{(i-1)} \gamma_{p:n,m},
\end{equation}
while $\tilde{\sigma}^{2(i)}_{x:{m,n}}$ are the \ac{sIC} conditional variances approximated by replacing the instantaneous values with the long-term statistics \cite{ParkerJSTSP2016,ItoTComm2023} as 
\begin{eqnarray}
\label{eq:variance_term_zeta_definition}
\hspace{-0.5ex}\tilde{\sigma}^{2(i)}_{x:{m,n}} \triangleq \mathbb{E}_{x,h} \big[ | \tilde{y}_{x:{m,n}}^{(i)} - \tilde{\gamma}_{x:{m,n}}^{(i)} x_m |^2 \big] && \nonumber \\
&&\hspace{-39.4ex} \approx \sum_{p=0}^P \hat{\sigma}^{2(i-1)}_{h:{n,p}} |\hat{y}^{(i)}_{x_m:n,p}|^2 + \sum_{q \neq m}^{N-1} \hat{\sigma}_{x:{n,q}}^{2(i-1)} |\tilde{\gamma}_{x:q,n}^{(i)}|^2 + N_0 + \\
&&\hspace{-37ex} \sum_{p=0}^P \hat{\sigma}^{2(i-1)}_{h:{n,p}} \sum_{q \neq m}^{N-1} \hat{\sigma}_{x:{n,q}}^{2(i-1)} |\gamma_{p:n,q}|^2\! +\! E_\mathrm{S} \sum_{p=0}^P \hat{\sigma}^{2(i-1)}_{h:{n,p}} |\gamma_{p:n,m}|^2 ,\nonumber
\end{eqnarray}
with $\hat{y}^{(i)}_{x_m:n,p}$ denoting the received signal estimate after cancellation of the $m$-th soft symbol estimate, which is given by
\begin{equation}
\label{eq:auxhvariancedatasoft}
\hat{y}^{(i)}_{x_m:n,p} \triangleq \sum_{q \neq m}^{N-1} \gamma_{p:n,q} \hat{x}_{n,q}^{(i-1)}.
\end{equation}

Similarly, the channel-centric replica can be computed as
\begin{equation}
\label{eq:soft_IC_for_channel_coeff}
\tilde{y}_{h:{p,n}}^{(i)} = y_n - \sum_{q \neq p}^P \hat{h}_{n,q}^{(i)} \cdot \hat{y}_{h:{n,q}}^{(i)},
\end{equation}
with the corresponding variance given by
\begin{eqnarray}
\label{eq:variance_term_zeta_definition_channel} 
\tilde{\sigma}^{2(i)}_{h:{p,n}}\! \triangleq \!\sum_{q \neq p}^P\! \hat{\sigma}_{h:{n,q}}^{2(i-1)} |\hat{y}_{h:{n,q}}^{(i)}|^2 \! + \!\!\! \sum_{m=0}^{N-1}\!\! \hat{\sigma}^{2(i-1)}_{x:{n,m}} |\tilde{\gamma}_{h_p:{m,n}}^{(i)}|^2  \!+\! N_0 +\!\! && \\
&&\hspace{-48ex} \sum_{q \neq p}^P\!\! \hat{\sigma}_{h:{n,q}}^{2(i)}\!\! \sum_{m=0}^{N-1}\!\! \hat{\sigma}^{2(i)}_{x:{n,m}} |\gamma_{q,nm}|^2 \!+ \!\sigma^2_h \sum_{m=0}^{N-1} \hat{\sigma}^{2(i)}_{x:{n,m}} |\gamma_{p:n,m}|^2 \!,\nonumber
\end{eqnarray}
where $\hat{y}_{h:{n,p}}^{(i)}$ is the channel-centric soft channel estimate and $\tilde{\gamma}_{h_p:{m,n}}^{(i)}$ is the corresponding \ac{sIC} effective channel gain of the $p$-th path, respectively given by
\begin{equation}
\label{eq:auxhvariancedata_and_channel}
\hat{y}_{h:{n,p}}^{(i)} \triangleq \sum_{m=0}^{N-1} \gamma_{p:n,m}\hat{x}_{n,m}^{(i-1)} \;\,\text{and}\;\,
\tilde{\gamma}_{h_p:{m,n}}^{(i)} \triangleq \sum_{q \neq p}^P \hat{h}_{n,q}^{(i-1)} \gamma_{q,nm}.
\end{equation}

\subsubsection{Belief Generation}

In order to generate the beliefs of all symbols, consider the effect of substituting equation \eqref{eq:elementwise_final_input_output_relation} into \eqref{eq:soft_IC_process} such that the  \ac{SGA} holds under the assumption that $N\times P$ is a sufficiently large number, and that the individual estimation errors in $\hat{h}_{n,p}^{(i-1)}$ and $\hat{x}_{n,q}^{(i-1)}$ are independent.

Then, as a result of \ac{SGA}, the belief corresponding to the $m$-th symbol $x_m$ at the $n$-th factor node is obtained by combining the contributions of all signals in $\mathbf{y}$, excluding $y_n$, under the \ac{PDF}
\vspace{-1ex}
\begin{equation}
\!\!\!\! p_{\text{x} | x_m} (\text{x} | x_m)\! =\! \prod_{q \neq n}^N\! p_{\text{y}_q | \text{x}_m}(\text{y}_q | x_m) \propto \text{exp} \bigg(\!\!\! - \!\frac{|\text{x} \!-\! \tilde{x}_{n,m}^{(i)}|^2}{\tilde{\sigma}_{\tilde{x}:{n,m}}^{2(i)}}\! \bigg),\!
\label{eq:PDF_extrinsic_belief}
\end{equation}
such that the desired beliefs and their variances are given by
\begin{equation}
\label{eq:mean_and_var_of_extrinsic_belief}
\tilde{x}_{n,m}^{(i)}\!\! \triangleq \!\tilde{\sigma}_{\tilde{x}:{n,m}}^{2(i)}\! \sum_{q \neq n}^N \!\!\frac{\tilde{\gamma}_{x:{m,q}}^{*(i)} \tilde{y}_{x:{m,q}}^{(i)}}{\tilde{\sigma}_{x:{m,q}}^{2(i)}}\;\,\text{and}\;\,
\tilde{\sigma}_{\tilde{x}:{n,m}}^{2(i)}\!\! \triangleq \!\!\bigg(\! \sum_{q \neq n}^N \!\!\frac{|\tilde{\gamma}_{x:{m,q}}^{(i)}|^2}{\tilde{\sigma}_{x:{m,q}}^{2(i)}} \bigg)^{\!\!\!-1}\!\!\!\!.
\end{equation}

Similarly, under \ac{SGA}, the extrinsic beliefs of the channel gains can be shown to follow the approximate distribution
\vspace{-1ex}
\begin{equation}
p_{\text{h} | h_p} (\text{h} | h_p) \propto \text{exp} \bigg( - \frac{|\text{h} - \!\tilde{\;\mathrm{h}}_{n,p}^{(i)}|^2}{\tilde{\sigma}_{\!\tilde{\;\mathrm{h}}:{n,p}}^{2(i)}} \bigg)   
\label{eq:PDF_extrinsic_belief_for_channel},
\end{equation}
with corresponding beliefs and variances given by

\quad\\[-5ex]
\begin{equation}
\label{eq:mean_and_variance_of_extrinsic_belief_for_channel}
\!\tilde{\;\mathrm{h}}_{n,p}^{(i)}\! \triangleq \!\tilde{\sigma}_{\!\tilde{\;\mathrm{h}}:{n,p}}^{2(i)} \sum_{q \neq n}^N \frac{\hat{y}_{h:{qp}}^{*(i)} \tilde{y}_{h:{p,q}}^{(i)}}{\tilde{\sigma}_{h:{p,q}}^{2(i)}}\;\,\text{and}\;\,
\tilde{\sigma}_{\!\tilde{\;\mathrm{h}}:{n,p}}^{2(i)}\! \triangleq \!\!\bigg(\! \sum_{q \neq n}^N \frac{|\hat{y}_{h:{qp}}^{(i)}|^2}{\tilde{\sigma}_{h:{p,q}}^{2(i)}} \bigg)^{\!\!-1}\!\!\!\!\!.
\end{equation}

\subsubsection{Soft RG}

Finally, the soft replicas of $x_m$ and $h_p$ can be obtained from the conditional expectation given the extrinsic beliefs, again thanks to the \ac{SGA} and under the assumption that the effective noise components in $\hat{x}_{n,m}^{(i)}, \forall (n,m)$ and $\hat{h}_{n,p}^{(i)}, \forall (n,p)$ are uncorrelated.
In particular, we have
\begin{equation}
\hat{x}_{n,m}^{(i)} =  \frac{\sum\limits_{x \in \mathcal{C}} x\cdot p_{\text{x} | x} \big(\text{x}|x;\tilde{x}_{n,m}^{(i)},\tilde{\sigma}_{\tilde{x}:{n,m}}^{2(i)}\big)\cdot p_{x}(x)}{\sum\limits_{x' \in \mathcal{C}} p_{\text{x} | x'} \big(\text{x}|x';\tilde{x}_{n,m}^{(i)},\tilde{\sigma}_{\tilde{x}:{n,m}}^{2(i)}\big)\cdot p_{x'}(x')},
\label{eq:Bayes_rule_data}
\end{equation}
\begin{equation}
\hat{h}_{n,p}^{(i)} = \frac{\int h\cdot p_{\text{h} | h} \big(\text{h}|h;\tilde{h}_{n,p}^{(i)},\tilde{\sigma}_{\tilde{h}:{n,p}}^{2(i)}\big)\cdot p_{h}(h)}{\int p_{\text{h} | h'} \big(\text{h}|h';\tilde{h}_{n,p}^{(i)},\tilde{\sigma}_{\tilde{h}:{n,p}}^{2(i)}\big)\cdot p_{h'}(h')},
\label{eq:Bayes_rule_channel}
\end{equation}
where $p_{\text{x} | x} \big(\text{x}|x;\tilde{x}_{n,m}^{(i)},\tilde{\sigma}_{\tilde{x}:{n,m}}^{2(i)}\big)$ and $p_{\text{h} | h} \big(\text{h}|h;\tilde{h}_{n,p}^{(i)},\tilde{\sigma}_{\tilde{h}:{n,p}}^{2(i)}\big)$ are the likelihood functions of the data and channel beliefs given in equations \eqref{eq:PDF_extrinsic_belief} and \eqref{eq:PDF_extrinsic_belief_for_channel}, respectively, only with the notation extended to explicitly show the parameters (means and variances) computed above.

\begin{algorithm}[t!]
  \caption{Proposed \ac{PBiGaBP}-based \ac{JCDE} Method}
  \label{alg:JCDE_PBiGaBP}
  \setlength{\baselineskip}{11pt}
  \textbf{Input:} receive signal vector $\mathbf{y}$, pilot symbol vector $\mathbf{x}_p$, delay-Doppler matrices $\mathbf{\Gamma}_p$, maximum number of iterations $i_\text{max}$, total constellation energy $E_\mathrm{S}$, noise power $N_0$, average channel power per path $\sigma^2_h$, and damping factors $\beta_x$ and $\beta_h$. \\
  \textbf{Output:} decoded symbols $\mathbf{x}_d$ and channel estimates $\hat{h}_p, \forall p$.
  \vspace{-2ex} 
  \begin{algorithmic}[1]  
  \STATEx \hspace{-3.5ex}\hrulefill
  \STATEx \hspace{-3.5ex}\textbf{Initialization}
  \STATEx \hspace{-3.5ex} - Set iteration counter to $i=0$ and amplitudes $c_x = \sqrt{E_\mathrm{S}/2}$
  \STATEx \hspace{-3.5ex} - Fix pilots to $\hat{x}_{n,m}^{(i)} = [\mathbf{x}_p]_m$ and set corresponding variances
  \STATEx \hspace{-2ex} to $\hat{\sigma}^{2(i)}_{x:{n,m}} = 0, \forall n,m \in \mathcal{M}_p$
  \STATEx \hspace{-3.5ex} - Set initial data estimates to $\hat{x}_{n,m}^{(0)} = 0$ and corresponding \STATEx \hspace{-2ex} variances to $\hat{\sigma}^{2(0)}_{x:{n,m}} = E_\mathrm{S}, \forall n,m \in \mathcal{M}_d$
  \STATEx \hspace{-3.5ex} - Set initial channel estimates to $\hat{h}_{n,p}^{(0)} = 0$ and corresponding
  \STATEx \hspace{-2ex} variances to $\hat{\sigma}^{2(0)}_{h:{n,p}} = \sigma_h^2, \forall n, p$
  \STATEx \hspace{-3.5ex}\hrulefill
  \STATEx \hspace{-3.5ex}\textbf{for} $i=1$ to $i_\text{max}$ \textbf{do}
  \STATEx \textbf{Channel Estimation}: $\forall n, p$
  \STATE Compute the variables $\hat{y}_{h:{n,p}}^{(i)}$ and $\tilde{\gamma}_{h_p:{m,n}}^{(i)}$ from eq.  \eqref{eq:auxhvariancedata_and_channel}.
  \vspace{-0.25ex}
  \STATE Compute soft signal $\tilde{y}_{h:{p,n}}^{(i)}$ from equation \eqref{eq:soft_IC_for_channel_coeff}.
  \vspace{-0.25ex}
  \STATE Compute soft signal variance $\tilde{\sigma}^{2(i)}_{h:{p,n}}$ from equation \eqref{eq:variance_term_zeta_definition_channel}.
  \vspace{-0.25ex}
  \STATE Compute extrinsic channel belief $\tilde{h}_{n,p}^{(i)}$ and its variance $\tilde{\sigma}_{\tilde{h}:{n,p}}^{2(i)}$ from equation \eqref{eq:mean_and_variance_of_extrinsic_belief_for_channel}.
  \vspace{-0.25ex}
  \STATE Compute denoised and damped channel estimate $\hat{h}_{n,p}^{(i)}$ from equations \eqref{eq:Gaussian_denoiser_channel} and \eqref{eq:Gaussian_denoiser_channel_damped}.
  \STATE Compute denoised and damped channel variance $\hat{\sigma}_{h:{n,p}}^{2(i)}$ from equations \eqref{eq:Gaussian_denoiser_MSE} and \eqref{eq:Gaussian_denoiser_MSE_damped}.
  \STATEx \textbf{Data Estimation}: $\forall n, m$
  \STATE Compute auxiliary variables $\tilde{\gamma}_{x:{m,n}}^{(i)}$ and $\hat{y}^{(i)}_{x_m:n,p}$ from eqs. \eqref{eq:effective_channel_gain_a_in_PDF} and \eqref{eq:auxhvariancedatasoft}, respectively.
  \STATE Compute soft signal $\tilde{y}_{x:{m,n}}^{(i)}$ from equation \eqref{eq:soft_IC_process}.
  \STATE Compute soft signal variance $\tilde{\sigma}^{2(i)}_{x:{m,n}}$ from equation \eqref{eq:variance_term_zeta_definition}.
  \STATE Compute extrinsic data belief $\tilde{x}_{n,m}^{(i)}$ and its variance $\tilde{\sigma}_{\tilde{x}:{n,m}}^{2(i)}$ from equation \eqref{eq:mean_and_var_of_extrinsic_belief}.
  \STATE Compute denoised and damped data estimates $\hat{x}_{n,m}^{(i)}$ from equations \eqref{eq:QPSK_denoiser} and \eqref{eq:QPSK_denoiser_damped}.
  \STATE Compute denoised and damped data estimate variances $\sigma_{x:{n,m}}^{2(i)}$ from equations \eqref{eq:MSE_x_m} and \eqref{eq:MSE_x_m_damped}.
  \STATEx \hspace{-3.5ex}\textbf{end for}
  \end{algorithmic}
\end{algorithm}

In order to compute the beliefs $\hat{x}_{n,m}^{(i)}$ with basis on equation \eqref{eq:Bayes_rule_data}, suffice it to notice that under the assumption that each element $x$ of the data vector is independently drawn with equal probability out of the $Q$-points of the constellation $\mathcal{C}$, the likelihood $p_x(x)$ is a Multinomial distribution of the $Q$-th order, such that for \ac{QPSK} modulation, we have
\begin{equation}
\hat{x}_{n,m}^{(i)}\! =\! c_x\! \cdot\! \bigg(\! \text{tanh}\!\bigg[ 2c_x \frac{\Real{\tilde{x}_{n,m}^{(i)}}}{\tilde{\sigma}_{\tilde{x}:{n,m}}^{2(i)}} \bigg]\!\! +\! j\text{tanh}\!\bigg[ 2c_x \frac{\Imag{\tilde{x}_{n,m}^{(i)}}}{\tilde{\sigma}_{\tilde{x}:{n,m}}^{2(i)}} \bigg]\!\bigg),\!\!
\label{eq:QPSK_denoiser}
\end{equation}
where $c_x \triangleq \sqrt{E_\mathrm{S}/2}$ represents the magnitude of the real and imaginary components of the \ac{QPSK} symbols.

After obtaining $\hat{x}_{n,m}^{(i)}$ as per equation \eqref{eq:QPSK_denoiser}, the final output is computed by damping the result with a damping factor $0 < \beta_x < 1$ in order to improve convergence \cite{Su_TSP_2015}, yielding
\begin{equation}
\label{eq:QPSK_denoiser_damped}
\hat{x}_{n,m}^{(i)} = \beta_x \hat{x}_{n,m}^{(i)} + (1 - \beta_x) \hat{x}_{n,m}^{(i-1)}.
\end{equation}

In turn, the corresponding variance $\hat{\sigma}^{2(i)}_{x:{n,m}}$ is first updated following equation \eqref{eq:MSE_x_m}, but with the updated symbol estimate, $i.e.$, $\hat{\sigma}^{2(i)}_{x:{n,m}} = E_\mathrm{S} - |\hat{x}_{n,m}^{(i)}|^2$, and then damped via
\begin{equation}
\label{eq:MSE_x_m_damped}
\hat{\sigma}^{2(i)}_{x:{n,m}} = \beta_x \hat{\sigma}_{x:{n,m}}^{2(i)} + (1-\beta_x) \hat{\sigma}_{x:{n,m}}^{2(i-1)}.
\end{equation}

Similarly to the above, in order to obtain the channel estimate belief, observe that the likelihood in equation \eqref{eq:Bayes_rule_channel} is a Gaussian-product distribution, with the mean of $p_{h}(h)$ equal to zero, which straightforwardly yields \cite{Julius_stanford_2011,Parker_TSP_2014}
\begin{equation}
\label{eq:Gaussian_denoiser_channel}
\hat{h}_{n,p}^{(i)} = \frac{\sigma^2_h \tilde{h}_{n,p}^{(i)}}{\tilde{\sigma}_{\tilde{h}:{n,p}}^{2(i)} + \sigma^2_h},
\end{equation}
and
\begin{equation}
\label{eq:Gaussian_denoiser_MSE}
\hat{\sigma}^{2(i)}_{h:{n,p}} = \frac{\sigma^2_h \tilde{\sigma}_{\tilde{h}:{n,p}}^{2(i)}}{\tilde{\sigma}_{\tilde{h}:{n,p}}^{2(i)} + \sigma^2_h},
\end{equation}
which are then damped with a damping factor $0 < \beta_h < 1$ as

\begin{equation}
\label{eq:Gaussian_denoiser_channel_damped}
\hat{h}_{n,p}^{(i)} = \beta_h \hat{h}_{n,p}^{(i)} + (1-\beta_h) \hat{h}_{n,p}^{(i-1)},
\end{equation}
and
\begin{equation}
\label{eq:Gaussian_denoiser_MSE_damped}
\hat{\sigma}_{h:{n,p}}^{2(i)} = \beta_h \hat{\sigma}_{h:{n,p}}^{2(i)} + (1-\beta_h) \hat{\sigma}_{h:{n,p}}^{2(i-1)}.
\end{equation}

The proposed \ac{PBiGaBP}-based \ac{JCDE} algorithm is summarized as a pseudocode in Algorithm \ref{alg:JCDE_PBiGaBP}.

\noindent \textit{Remark 1:} Although the proposed \ac{PBiGaBP}-based \ac{JCDE} algorithm is designed based on the Bayes-optimal \ac{GaBP} framework, to the best of our knowledge, there exists no asymptotic analysis of the large-system limit for iterative convergence analysis of Bayesian parametric bilinear inference algorithms.
While semi-analytical methods have been proposed for the prediction of the number of iterations required for fixed-point convergence in large scale systems \cite{Akrout_ASILOMAR_2022,YuanTSP2021}, they cannot be used under correlated observations.
Therefore, for the performance analysis detailed in Section \ref{sec:JCDE_Simulations}, rigorous numerical simulations are provided to demonstrate the superiority of the proposed technique.

\vspace{-2ex}
\subsection{Complexity Analysis}
\label{sec:JCDE_Complexity}

Since there are no high-complexity matrix inversions in the proposed \ac{PBiGaBP}-based \ac{JCDE} scheme detailed in Algorithm \ref{alg:JCDE_PBiGaBP}, the order of complexity strictly depends on the scalar operations executed.
Therefore, the computational complexity of the proposed method is $\mathcal{O}(N^2(P+1))$ per iteration, which is linear in terms of the operations performed, and of a much lower complexity than \ac{SotA} methods based on \ac{LMMSE} filtering, which incur a cost of $\mathcal{O}(N^3)$ due to a computation of \ac{LMMSE} filter involving a matrix inversion operation.
%

The proposed method also has comparable complexity to the \ac{SotA} \ac{PBiGAMP} scheme given in \cite{Furuta_CCNC_2024}, which also is of order $\mathcal{O}(N^2(P+1))$, but far outperforms the latter method in terms of performance, therefore offering an excellent performance-complexity tradeoff.

\vspace{-2ex}
\subsection{Performance Analysis}
\label{sec:JCDE_Simulations}

In this subsection, we evaluate the performance of the proposed \ac{PBiGaBP} algorithm at the \ac{UE}, under \ac{OFDM}, \ac{OTFS} and \ac{AFDM} waveforms.
To that end, we consider the scenario depicted in Figure \ref{fig:ISAC_system_model} with $1$ \ac{LoS} path and $4$ \ac{NLoS} paths ($i.e.,$ $P_\mathrm{U}=4$) seen at the \ac{UE}; and assume a \ac{mmWave} system\footnote{Although the proposed scheme can be used for arbitrary frequencies and bandwidths, our focus here is on \ac{mmWave} systems due to the trend towards moving to higher frequencies in 6G and beyond \cite{WymeerschPIMRC2021}.} operating at $70 \, \text{GHz}$ with a bandwidth of $20 \, \text{MHz}${\footnote{With $f_\mathrm{S}$ set to be equal to the bandwidth.}} and employing \ac{QPSK} symbols.
For the channel model, we consider a maximum normalized delay index of $20$ and a maximum normalized digital Doppler shift index of $0.25$ (i.e., $\ell_\text{max} \triangleq \tau_\text{max} f_\mathrm{S} = 20$ and $f_\text{max} \triangleq \frac{N \nu_\text{max}}{f_\mathrm{S}} = 0.25$), which allows for a maximum unambiguous range of $75 \, \text{m}$ and a maximum unambiguous velocity of $602 \, \text{km/h}$, respectively. 
This allows for the random generation of the path delays $\tau_p$ using a uniform distribution across $[0,\tau_\text{max}]$ and the use of Jakes Doppler spectrum \cite{Bemani_TWC_2023} for the path Doppler shifts as $\nu_p = \nu_\text{max} \cos(\theta_p)$, where $\theta_p$ is uniformly distributed over $[-\pi,\pi]$ at each Monte Carlo iteration.
As for frame parameters, we consider $K = B = 32$ and $M=4$ for \ac{OTFS}, and in order to sustain a fair comparison, $N=128$ for \ac{OFDM}\footnote{While there exists a multitude of ways to distribute the time-frequency resources for \ac{OFDM} systems \cite{Gaudio_TWC_2022}, we choose to use the classic \ac{OFDM} structure with symbols spread over the entire time period for applicability with the utilized doubly-dispersive channel structure.} and \ac{AFDM}. 
Finally, regarding the \ac{PBiGaBP} algorithm, we set $\beta_x, \beta_h = 0.3$, the maximum number of iterations $i_\text{max} = 40$, the constellation power $E_\mathrm{S}=1$ and the average channel power per path $\sigma_h^2 = 1$.

\begin{figure}[t!]
  \centering
  \captionsetup[subfloat]{labelfont=small,textfont=small}
  \subfloat[BER Performance.]{{\includegraphics[width=0.95\columnwidth]{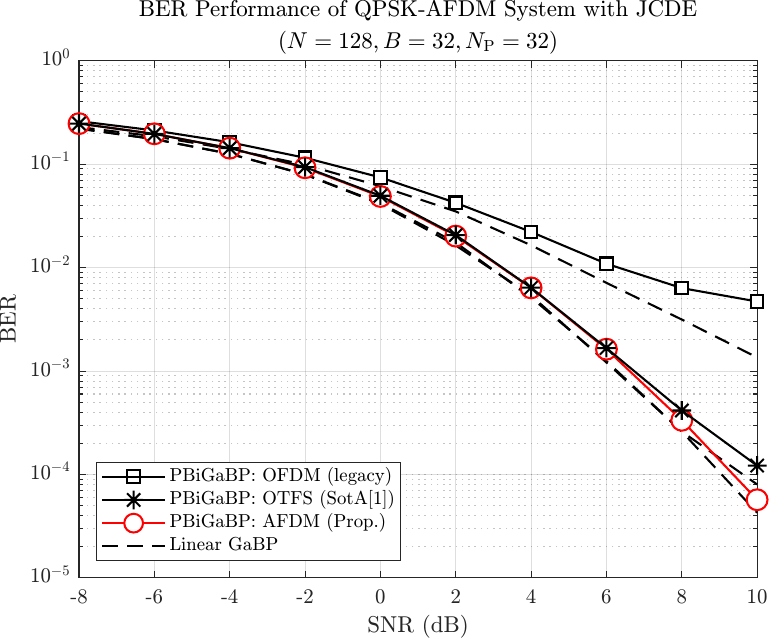}}}%
  \label{fig:BER_plot}
  \subfloat[NMSE Performance.]{{\includegraphics[width=0.95\columnwidth]{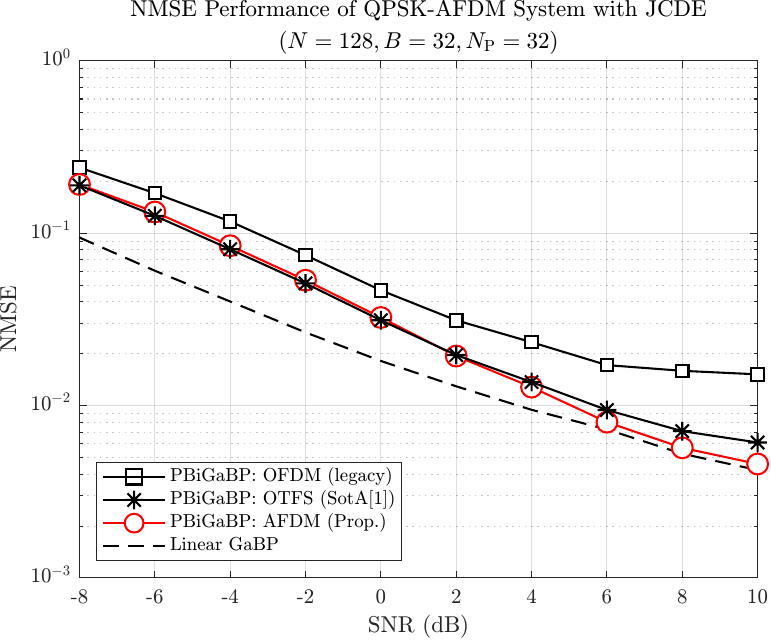}}}%
  \label{fig:NMSE_plot}
  \caption{Performance of OFDM, OTFS and AFDM at the UE via the proposed PBiGaBP algorithm compared to the SotA methods.}
  \label{fig:XXXM_Performance_plot}
\end{figure}

First, using the Linear \ac{GaBP}\footnote{Linear \ac{GaBP} is a technique that uses \ac{GaBP} for both channel estimation (under full knowledge of the frame) and data decoding (under full knowledge of the channel coefficients), combined with the use of an \ac{LMMSE} filtering-based initialization for channel estimation. The approach provides a lower bound on the performance for the proposed method in exchange for a large computational cost.} as a baseline, we compare in Figure \ref{fig:XXXM_Performance_plot} the communications and channel estimation performances, in terms of \ac{BER} and \ac{NMSE}, respectively, achieved by \ac{PBiGaBP} employed over \ac{OTFS}, \ac{OFDM} and \ac{AFDM} waveforms. 
In this comparison, all systems
employ an equal number $N_p$ of pilot symbols, although each system has its own structure (see Section \ref{sec:SystemModel} and Figure \ref{fig:ISAC_frame_design}).

The results show that \ac{PBiGaBP}-based \ac{JCDE} with \ac{AFDM} and \ac{OTFS} outperforms a system employing the same proposed technique but employing a legacy \ac{OFDM} waveform, both in terms of \ac{BER} and \ac{NMSE}, with a slight advantage of \ac{AFDM} over the \ac{OTFS} alternative.

\begin{figure}[t]
  \centering
  \captionsetup[subfloat]{labelfont=small,textfont=small}
  \subfloat[BER Performance.]{{\includegraphics[width=0.95\columnwidth]{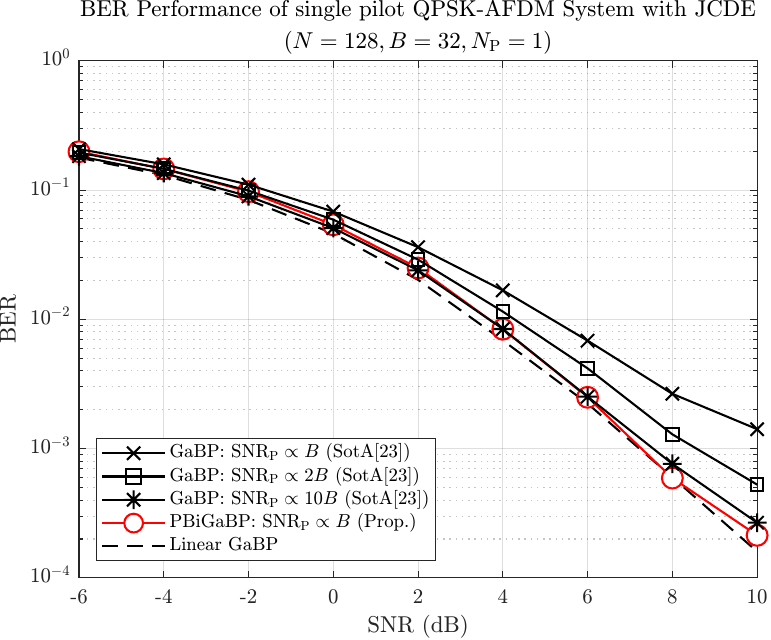}}}%
  \label{fig:BER_A_plot}
  \subfloat[NMSE Performance.]{{\includegraphics[width=0.95\columnwidth]{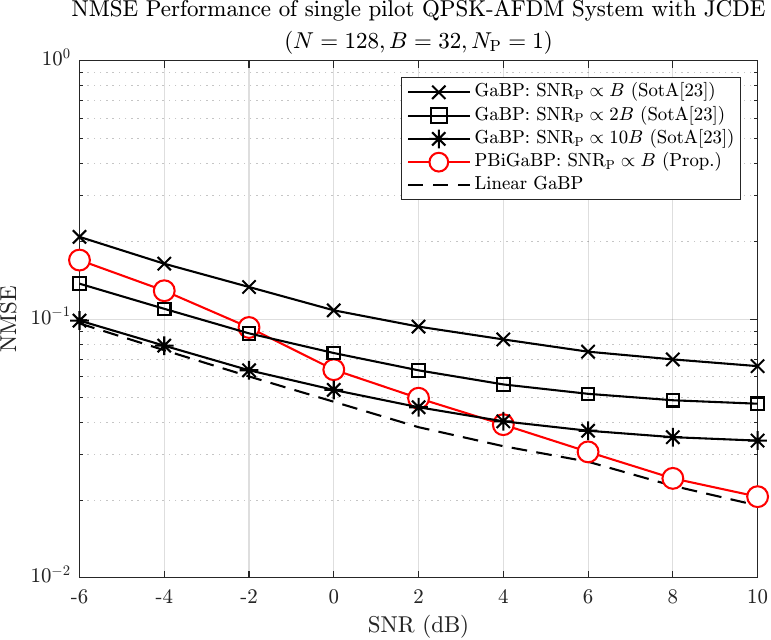}}}%
  \label{fig:NMSE_A_plot}
  \vspace{-0.5ex}
  \caption{Performance of the single pilot QPSK-AFDM system at the UE via the proposed PBiGaBP algorithm compared to the SotA methods.}
  \label{fig:BER_AFDM_plot}
  \vspace{-2ex}
\end{figure}

We highlight that a comparison between the proposed \ac{JCDE} algorithm and other techniques from related literature is not offered in Figure \ref{fig:XXXM_Performance_plot}, as we are \ul{unaware} of any alternative \ac{JCDE} solution for \ac{AFDM} systems.
Therefore, Figure \ref{fig:BER_AFDM_plot} compares the proposed \ac{PBiGaBP}-based \ac{JCDE} both in terms of \ac{BER} and \ac{NMSE} performance against the \ac{GaBP} alternative to the \ac{ML}-based \ac{AFDM} channel estimation method from \cite{Bemani_WCL_2024}, where channel estimation is done solely based on the single pilot via \ac{GaBP} with an accompanying \ac{GaBP}-based data decoding.

We emphasize, however, that the technique in \cite{Bemani_WCL_2024} is \ul{not} a \ac{JCDE} scheme, being rather strictly pilot-based like traditional channel estimation, such that in comparison, the latter \ac{SotA} method enjoys the advantage of not dealing with the additional burden of performing data detection, which the proposed algorithm does.
In addition, as indicated in the legend of the figure, the comparison awards the \ac{SotA} approach a power advantage by allowing the latter to employ pilot signals with higher power than those used in the proposed method.

The results show that the proposed \ac{PBiGaBP}-based \ac{JCDE} scheme outperforms the traditional pilot-only \ac{GaBP}-based method for the single pilot \ac{AFDM} frame structure proposed in \cite{Bemani_WCL_2024} and illustrated in Figure \ref{fig:ISAC_frame_design}, even when the latter employs a pilot with 10 times as higher power, which is a consequence of the fact that the proposed scheme takes advantage not only of the pilot, but also payload symbols ($i.e.,$ \textit{soft} pilots) to estimate the channel.
The results also motivates us hereafter to no longer consider \ac{OFDM}, as it is clear that it is outperformed by \ac{OTFS} and \ac{AFDM} in doubly-dispersive channels.

\section{Radar Parameter Estimation in Doubly Dispersive Channels}
\label{sec:Generalized_Radar_Parameter_Estimation_at_BS}

Having demonstrated the effectiveness of our proposed \ac{PBiGaBP}-based \ac{JCDE} scheme for \ac{AFDM}, we next consider the radar parameter estimation problem\footnote{For the sake of simplicity, we concentrate here on the parameter estimation -- as opposed to target acquisition/detection -- part of the radar problem, such that the total number of targets is assumed to be known \cite{Gaudio_RadarConf20_2020}, and the target association problem is assumed to be solvable \cite{Bar-ShalomCSM2009}. It is easy to perceive from the approach introduced, however, that an extension of the work here presented to joint detection and parameter estimation, such as performed in \cite{Gaudio_RadarConf20_2020}, is possible. That will, however, be pursued in a follow-up work.}, also referred to as \ac{DD} domain channel estimation problem \cite{FishTIT2013}.
To that end, a typical \ac{DD} channel \cite{Mehrotra_TCom_2023} is hereafter assumed, characterized by a maximum normalized delay spread $\text{max}(\ell_p)$ and a corresponding digital normalized Doppler spread $\text{max}(f_p)$, satisfying the relations $\text{max}(\ell_p) << N$ and $\text{max}(f_p) << N$.

Let us start by pointing out that the channel impulse response given in equation \eqref{eq:doubly_dispersive_time_delay_channel} can also be represented in the \ac{DD} domain \cite{Rou_SPM_2024} as 
\vspace{-1ex}
\begin{equation}
\label{eq:DD_channel_definition}
h(\tau,\nu) = \sum_{p=0}^P h_p \delta(\tau - \tau_p) \delta(\nu - \nu_p),
\vspace{-0.5ex}
\end{equation}
which for a finite number $P$ of paths, and under the aforementioned bounding assumption on delay and Doppler parameters, can be rewritten as
\begin{equation}
\label{eq:DD_channel_definition_frac_Dopp}
h(\tau,\nu) = \sum_{k=0}^{K_\tau - 1} \sum_{d=0}^{D_\nu - 1} h_{k,d} \delta(\tau - \tau_k) \delta(\nu - \nu_d),
\end{equation}
where $K_\tau$ and $D_\nu$ are numbers large enough to define a sufficiently fine grid discretizing the region determined by the maximum delay and Doppler values.

Notice that if the resolution of such a grid is sufficiently fine, the only non-zero terms of the sum in equation \eqref{eq:DD_channel_definition_frac_Dopp} are those where both $\tau_k\approx \tau_p$ and $\nu_d \approx \nu_p$, which in turn implies that estimating these radar parameters amounts to estimating the $P$ channel gains such that $h_{k,d}\neq 0$.
This suggests that the radar parameter estimation problem can be formulated as a canonical sparse signal recovery problem \cite{GongSPAWC2023} as follows.

First, rewriting the time-continuous model of equation \eqref{eq:DD_channel_definition_frac_Dopp} in terms of the sampled equivalent in matrix form yields
\begin{equation}
\label{eq:TD_channel_w_frac_Dopp_matrix}
\mathbf{\Bar{H}} \triangleq \sum_{k=0}^{K_\tau - 1} \sum_{d=0}^{D_\nu - 1} h_{k,d} \cdot \mathbf{\Phi}_k \cdot \boldsymbol{\Omega}^{f_d} \cdot \mathbf{\Pi}^{\ell_k} \in \mathbb{C}^{N\times N},
\end{equation}
where we highlight that the model described by equation \eqref{eq:TD_channel_w_frac_Dopp_matrix} is identical to the equivalent channel implicitly defined in equation \eqref{eq:channel_matrix_TD_general}, only with the summation rewritten over the \ac{DD} grid, as opposed to the channel paths.

\begin{subequations}
\label{eq:sparse_RPE}
Then, the generalized receive signal model given in equation \eqref{eq:final_generalized_input_output_relation} can be rewritten as 
\vspace{-1ex}
\begin{equation}
\label{eq:modified_input_output_relation_for_estimation}
\mathbf{y} \!=\!\! \sum_{k=0}^{K_\tau\! -\! 1} \sum_{d=0}^{D_\nu - 1}\!\! \underbrace{\mathbf{\Gamma}_{k,d} \cdot \mathbf{x}}_{\mathbf{e}_{k,d}\in \mathbb{C}^{N\times 1}}\!\!\! \cdot h_{k,d} + \tilde{\mathbf{w}} = \mathbf{E} \cdot \mathbf{h} + \tilde{\mathbf{w}} \in \mathbb{C}^{N\times 1},
\vspace{-1.5ex}
\end{equation}
where we implicitly defined the vectors $\mathbf{e}_{k,d}\triangleq\mathbf{\Gamma}_{k,d} \cdot \mathbf{x}$, and explicitly identified the dictionary matrix $\mathbf{E} \in \mathbb{C}^{N\times K_\tau D_\nu}$, and sparse channel vector $\mathbf{h} \in \mathbb{C}^{K_\tau D_\nu \times 1}$, respectively defined as
%
\begin{equation}
\label{eq:sparse_dict_matrix_def}
\mathbf{E}\! \triangleq\! [\mathbf{e}_{0,0}, \dots, \mathbf{e}_{0,D_\nu\! -\! 1}, \dots, \mathbf{e}_{K_\tau\! -\! 1,0}, \dots, \mathbf{e}_{K_\tau\! -\! 1,D_\nu \!-\! 1}],
\vspace{-1ex}
\end{equation}
\begin{equation}
\label{eq:sparse_channel_vector}
\mathbf{h} \!\triangleq\! [h_{0,0}, \dots, h_{0,D_\nu\! -\! 1}, \dots, h_{K_\tau\! -\! 1,0}, \dots, h_{K_\tau\! -\! 1,D_\nu\! -\! 1}]\trans\!\!\!.
\end{equation}
\end{subequations}

In summary, the problem of estimating the set of radar parameters $\{\tau_p,\nu_p\}, \forall p$ reduces to estimating the sparse channel vector $\mathbf{h}$, with the delays and Doppler shifts obtained from the corresponding indices $(k,d)$ where $h_{k,d}\neq 0$, given the received signal $\mathbf{y}$ and the dictionary matrix $\mathbf{E}$.
As a consequence of the approach, the assumption that $P$ is known is relaxed into the assumption that the channel paths are orthogonal in the delay-Doppler grid, such that $P$ can be unambiguously inferred from $\mathbf{h}$.
Finally, the method also implies the assumption that the transmit vector $\mathbf{x}$ is known at the radar receiver, which in turn translates to a number of possible scenarios \cite{Rou_SPM_2024}, including those described below.

\emph{Monostatic \ac{RPE}}: A first scenario is the classic monostatic case, in which the \ac{BS} processes round-trip signals reflected by surrounding scatters (targets) in order to extract their distances and velocities.
Here, a main practical challenge is the self-interference caused by signals transmitted by the \ac{BS} at the same time when reflected signals return.
The issue is often ignored in related literature on \ac{OFDM}- and \ac{OTFS}-based monostatic \ac{RPE} \cite{GaudioTWC2020, Ni_ISWCS_2022}, implying a strong and not entirely practical assumption that self-interference is perfectly avoided by \ac{FD} technologies.
To emphasize this point, it has been shown that \ac{AFDM} has an advantage over \ac{OFDM} and \ac{OTFS} as it allows for the concentration of a single pilot symbol within a transmission block\footnote{Observe from equations \eqref{eq:DAF_input_output_relation} and \eqref{eq:DAF_domain_effective_channel} that unlike \ac{OFDM} and \ac{OTFS}, the \ac{AFDM} receive signal occupies the entire time-frequency span even when a single entry in the transmit vector $\mathbf{x}$ is non-zero, such that concentrating piloting to a single symbol does not sacrifice the \ac{DoF} of the estimator.}, as illustrated in Figure \ref{fig:ISAC_frame_design}, with the corresponding self-interference mitigated via a simple analog dechirping and filtering operation \cite{Bemani_WCL_2024}.
We will therefore focus on this case.

\emph{User-centric \ac{RPE}}: In this case, the \ac{BS} first transmits a pilot symbol vector used by the \ac{UE} to perform \ac{RPE}, and subsequently transmits data from which the \ac{UE} proceeds to perform \ac{JCDE} as described in Section \ref{sec:Generalized_JCDE_at_UE} in order to track the channel and decode the information.
Under the assumption that delay and Doppler shifts change at a rate slower than $h_p$ \cite{MuraliITA2018}, the alternation of \ac{JCDE} and \ac{RPE} stages would enable the system to minimize the overhead associated with frequent pilot transmissions.
Due to the lack of space, this scenario will not be further developed here, but we have shown in \cite{TakahashiTWC_JCDE2024} that a well-designed \ac{JCDE} algorithm is capable of tracking time-varying channels even without periodic pilot transmissions, as long as the \ac{BER} of the system is maintained sufficiently low, $e.g.$ via adequate channel coding.
It is not difficult to envision that by combining the contribution of this article with the one in \cite{TakahashiTWC_JCDE2024}, the user-centric \ac{RPE} scenario can be addressed.
%

\emph{Bistatic \ac{RPE}}: This refers to the case when passive receivers -- could be \acp{BS} or \acp{UE} -- which are placed at different locations with respect to the actively transmitting \ac{BS} perform \ac{RPE} using the received signals from the active \ac{BS} \cite{KanhereVTC2021,LeyvaVTC2022,ZhuWCL2024,Ranasinghe_WiOpt_2024}.
The scenario can be similar either to the latter user-centric case, if the passive receiver have no knowledge of the transmit vector $\mathbf{x}$ \cite{Ranasinghe_WiOpt_2024}, or to the monostatic case, if $\mathbf{x}$ can be assumed known via pilot symbols, a preamble or a reference signal \cite{KanhereVTC2021,LeyvaVTC2022,ZhuWCL2024} to all the passive receivers.
It is clear, therefore, that this scenario is a generalization of both, which is best addressed in a dedicated follow-up article, especially since the latter assumption -- if \acp{BS} are considered -- requires them to be interconnected via a fronthaul network, which in turn might be subjected to impairments such as delays, noise, capacity limitations etc.

For the rest of this manuscript, \ac{wlg} and due to space limitations, we will focus on monostatic \ac{RPE}.

\vspace{-3ex}
\subsection{Reference SotA Method: SBL via Expectation Maximization}
\label{sec:SotA_SBL_via_EM}

Before we introduce our contribution, let us briefly review an effective \ac{SotA} channel estimation approach, which is here adapted to solve the \ac{RPE} problem, and which will be used as a reference for performance assessment.
The technique, which is based on the \ac{SBL} framework, was proposed in \cite{Mehrotra_TCom_2023} for \ac{OTFS} systems, but can also be employed to \ac{AFDM} thanks to the unified model of Section \ref{sec:SystemModel} and the resulting sparse linear representation of equation \eqref{eq:sparse_RPE}.
To see this, recall that the \ac{SBL} approach models the sparse channel vector $\mathbf{h}$ as a parameterized Gaussian distribution of the form
\begin{equation}
f(\mathbf{h};\mathbf{\Xi}) = \prod_{m=0}^{K_\tau D_\nu - 1} \frac{1}{\pi \xi_m} \text{exp} \bigg( - \frac{|h_m|^2}{\xi_m} \bigg),
\end{equation}
where $\xi_m$ denotes the unknown hyperparameter corresponding to the $m$-th component of the vector $\mathbf{h}$, collected into the matrix $\mathbf{\Xi} \triangleq \text{diag} \big( \{ \xi_m \}_{m=0}^{K_\tau D_\nu - 1} \big) \in \mathbb{R}_+^{K_\tau D_\nu \times K_\tau D_\nu}$.

Then, estimates $\mathbf{\!\hat{\;h}}$ of the sparse vector and its corresponding hyperparameter matrix $\hat{\mathbf{\Xi}}$ are computed iteratively leveraging the \ac{EM} method, which requires knowledge of the noise covariance matrix associated with the received signal model.

For the case of \ac{OTFS}, from the definition of $\tilde{\mathbf{w}}_\text{OTFS}$ given just after equation \eqref{eq:DD_input_output_relation} and under the assumption of rectangular pulse-shaping, we readily have
\begin{eqnarray*}
\mathbf{R_w} \!\triangleq \!\mathbb{E}\Big[\tilde{\mathbf{w}}_\text{OTFS}\!\cdot\! \tilde{\mathbf{w}}_\text{OTFS}\herm \Big]\!\!=\!\mathbb{E}\Big[\big[(\mathbf{F}_M\! \otimes\! \mathbf{I}_K)\mathbf{w} \big]\!\! \cdot\!\! \big[ (\mathbf{F}_M \!\otimes\! \mathbf{I}_K) \mathbf{w} \big]\herm\Big]&&
\nonumber \\
&&\hspace{-58ex}= (\mathbf{F}_M\! \otimes\! \mathbf{I}_K)\! \cdot\! N_0 \mathbf{I}_{N}\! \cdot\! (\mathbf{F}_M\! \otimes\! \mathbf{I}_K)\herm\! =\! N_0 \Big[ \mathbf{F}_M \mathbf{F}_M\herm\! \otimes\! \mathbf{I}_K \mathbf{I}_K\herm\Big]
\end{eqnarray*}
\vspace{-3ex}
\begin{equation}
\hspace{-15.3ex}= N_0 \Big[ \mathbf{I}_M \otimes \mathbf{I}_K \mathbf{I}_K\herm\Big]
= N_0 \mathbf{I}_N \in \mathbb{C}^{N \times N}.
\label{eq:OTFS_covariance_matrix_comp}
\end{equation}

\vspace{-1ex}
\begin{algorithm}[H]
\caption{\ac{SBL}-based Radar Parameter Estimation \cite{Mehrotra_TCom_2023}}
\label{alg:SBL-EM}
\textbf{Input:} Receive signal $\mathbf{y}$, dictionary matrix $\mathbf{E}$, noise covariance matrix $\mathbf{R_w}$, and stopping criteria parameters $\epsilon$ and $i_\text{max}$.\\
\textbf{Output:} Sparse channel estimate vector $\mathbf{\!\hat{\;h}}$.\\
\textbf{Initialization:} Set counter to $i = 0$, and initial hyperparameters to $\hat{\mathbf{\Xi}}^{(0)} = \mathbf{I}_{K_\tau D_\nu}$ and $\hat{\mathbf{\Xi}}^{(-1)} = \mathbf{0},$
\vspace{0.2ex}
\hrule
\begin{algorithmic}[1]

\STATEx \hspace{-4ex}{\textbf{while}} {$|| \hat{\mathbf{\Xi}}^{(i)} - \hat{\mathbf{\Xi}}^{(i-1)} ||^2_F > \epsilon$ and $i < i_\text{max}$} {\textbf{do}}
\STATE $i \leftarrow i + 1$
\STATEx \hspace{-3ex} \textbf{E-step:} Compute \textit{aposteriori} covariance and estimate vector
\vspace{-2ex}
\STATE \hspace{8ex} $\bm{\Sigma}^{(i)} = \big[ \mathbf{E}\herm \mathbf{R}^{-1}_\mathbf{w} \mathbf{E} + (\hat{\mathbf{\Xi}}^{(i-1)})^{-1} \big]^{-1}$,
\vspace{0.5ex}
\STATE \hspace{8ex} $\mathbf{\!\hat{\;h}}^{\!(i)} = \bm{\Sigma}^{(i)} \mathbf{E}\herm \mathbf{R}_\mathbf{w}^{-1} \mathbf{y} $.
\vspace{0.5ex}
\STATEx \hspace{-3ex} \textbf{M-step:} Update the hyperparameters $\hat{\mathbf{\Xi}}^{(i)}$, with
\vspace{1ex}
\STATE \hspace{8ex}  $\hat{\xi}_m^{(i)} = \sigma_{m,m}^{(i)} + \big|\hat{h}_m^{(i)}\big|^2\!\!,\;\forall\,m$.
\vspace{0.5ex}
\STATEx \hspace{-4ex}{\textbf{end}} {\textbf{while}}

\end{algorithmic}
\label{SBL-EM}
\end{algorithm}

In turn, taking into account the definition of $\tilde{\mathbf{w}}_\text{AFDM}$ given after equation \eqref{eq:DAF_input_output_relation}, we have for the \ac{AFDM} case
\begin{eqnarray*}
\mathbf{R_w} \!\triangleq \!\mathbb{E}\Big[\tilde{\mathbf{w}}_\text{AFDM}\!\cdot\! \tilde{\mathbf{w}}_\text{AFDM}\herm \Big]\!\!=\!\mathbb{E}\Big[ \! \big[ (\!\mathbf{\Lambda}_2 \mathbf{F}_{\!N} \mathbf{\Lambda}_1 \!) \mathbf{w} \big]\!\! \cdot \!\!\big[ (\!\mathbf{\Lambda}_2 \mathbf{F}_{\!N} \mathbf{\Lambda}_1 \!) \mathbf{w} \big]\herm \!\Big]&&
\nonumber \\
&&\hspace{-58ex}= \mathbb{E}\Big[\big[ (\!\mathbf{\Lambda}_2 \mathbf{F}_{\!N} \mathbf{\Lambda}_1 \!) \mathbf{w} \big]\! \cdot\! \big[ \mathbf{w}\herm (\!\mathbf{\Lambda}_2 \mathbf{F}_{\!N} \mathbf{\Lambda}_1 \!)\herm \big]\Big]
\nonumber \\
&&\hspace{-58ex}= (\!\mathbf{\Lambda}_2 \mathbf{F}_{\!N} \mathbf{\Lambda}_1 \!)\! \cdot\! N_0 \mathbf{I}_{N}\! \cdot\! (\!\mathbf{\Lambda}_2 \mathbf{F}_{\!N} \mathbf{\Lambda}_1 \!)\herm\! =\! N_0 \mathbf{I}_N \in \mathbb{C}^{N \times N}.
\end{eqnarray*}
\vspace{-6.5ex}
\begin{equation}
\label{eq:AFDM_covariance_matrix_comp}
\end{equation}

A pseudocode summary of the \ac{SBL}-based \ac{RPE} scheme is offered in Algorithm \ref{SBL-EM}.

\subsection{Proposed Method: PDA-based Message Passing Estimator}
\label{sec:Proposed_PDA_MP}

In this section, we propose a novel \ac{PDA}-based \ac{RPE} scheme designed under the assumption that the prior distributions of the elements $\hat{h}_m$ of the sparse channel vector estimates $\mathbf{\!\hat{\;h}}$ are of Bernoulli-Gaussian type.
In other words, we model the unknown channel to be estimated at the $i$-th iteration of the proposed scheme as 
\vspace{-0.5ex}
\begin{equation}
\label{eq:h_m_estimate_dist}
\hat{h}_m^{(i)} \sim p_{\,\text{h}_m} (\text{h}_m;\bm{\theta}^{(i)}),
\vspace{-2ex}
\end{equation}
with
\vspace{-0.5ex}
\begin{equation}
\label{eq:h_m_pdf}
\!\!p_{\,\text{h}_m} (\text{h}_m;\bm{\theta}^{(i)}) \triangleq\! (1\! -\! {\rho}^{(i)}) \delta(h_m)\! + \!{\rho}^{(i)} \mathcal{CN}\big(h_m;{\bar{h}}^{(i)},{\bar{\sigma}}^{(i)} \big),\!\!
\vspace{-0.5ex}
\end{equation}
where $\bm{\theta}^{(i)} \triangleq [{\rho}^{(i)}, {\bar{h}}^{(i)},{\bar{\sigma}}^{(i)}]$ carries all three parameters of the distribution, namely, sparsity rate, mean and variance.

As implied by the notation, the parameter vector $\bm{\theta}^{(i)}$ must be updated iteratively, which in similarity to the \ac{SBL} approach of \cite{Mehrotra_TCom_2023} can be accomplished via the \ac{EM} algorithm.
Unlike the aforementioned \ac{SotA} method, however, the estimates of $\mathbf{h}$ in the proposed scheme are obtained via a message-passing algorithm, which is described in the sequel.

For later convenience, we define the soft replica (i.e., tentative estimate) of $h_m$ as $\big\{ \hat{h}_{m}^{(i)} \big\} $ such that its \ac{MSE} can be expressed as
\vspace{-0.5ex}
\begin{equation}
\hat{\sigma}_{h:m}^{2(i)} \triangleq \mathbb{E}_{\text{h}_m} \big[  |h_m - \hat{h}_{m}^{(i)}|^2   \big].
\label{eq:MSE_h_PDA}
\end{equation}

Following steps similar to those in Section \ref{sec:SotA_GAMP_GaBP}, we have

\subsubsection{Soft IC}

The \ac{sIC} expression corresponding to an estimate of $h_m$, is given by
\vspace{-2ex}
\begin{equation}
\label{eq:Soft_IC_PDA}
\tilde{\mathbf{y}}_{h:m}^{(i)}\! = \mathbf{y}\! -\!\!\! \sum_{q\neq m}\!\! \mathbf{e}_{q} \hat{h}_{q}^{(i)}  \!=\! \mathbf{e}_{m}h_m \!+\!\!\!\!\!\!\! \overbrace{\sum_{q \neq m}^{K_\tau D_\nu}  (\mathbf{e}_{q} h_q\! -\! \mathbf{e}_{q} \hat{h}_{q}^{(i)})\! +\! \tilde{\mathbf{w}}}^{\text{residual interference+noise component}}\!\!\!, \!\!
\end{equation}
where $\mathbf{e}_m$ is the $m$-th column of the dictionary matrix $\mathbf{E}$.

It follows from the \ac{CLT} that, under large-system conditions, the residual interference-plus-noise component can be approximated by a multivariate complex Gaussian variate.
In other words, the \ac{VGA} can be applied such that the conditional \ac{PDF} of the beliefs $\tilde{\mathbf{y}}_{h:m}^{(i)}$, given $h_m$, can be expressed as
\vspace{-0.5ex}
\begin{equation}
\label{eq:VGA_y_given_h}
\!\!\tilde{\mathbf{y}}_{h:m}^{(i)}\!\!\sim\! p_{\textbf{y} | \text{h}_{m}} \!(\textbf{y} | h_{m}\!) 
\!\propto\! \text{exp} \!\Big[\! -\! \big(\! \textbf{y}\! -\! \mathbf{e}_{m} h_{m}\! \big)\herm \mathbf{\Sigma}^{-1(i)}_{m}\! \big(\! \textbf{y} \!- \!\mathbf{e}_{m} h_{m} \!\big)\! \Big]\!,\!\!\!
\end{equation}
where $\textbf{y}$ is an auxiliary variable, and the conditional covariance matrix $\mathbf{\Sigma}_m^{(i)}$ is given by
\vspace{-1ex}
\begin{align}
\mathbf{\Sigma}_m^{(i)} & \triangleq \mathbb{E}_{\textbf{h,$\tilde{\mathbf{w}}$}|\hat{h}_m \neq h_m} \bigg[ \big( \tilde{\mathbf{y}}_{h:m}^{(i)} - \mathbf{e}_{m} h_{m} \big) \big( \tilde{\mathbf{y}}_{h:m}^{(i)} - \mathbf{e}_{m} h_{m} \big)\herm  \bigg]
\nonumber \\[-1ex]
& = \sum_{q \neq m}^{K_\tau D_\nu} \hat{\sigma}_{h:q}^{2(i)} \mathbf{e}_q \mathbf{e}_q\herm + N_0 \mathbf{I}_N,
\label{eq:covariance_matrix_PDA}
\end{align}
with $N_0$ denoting the noise power.

\subsubsection{Belief Generation}
The beliefs associated with the estimate of the $m$-th channel entry $h_m$ can be obtained by combining the contributions of all \ac{sIC} beliefs $\tilde{\mathbf{y}}_{h:m}^{(i)}$, under the PDF
\begin{equation}
p_{\text{h} | h_m} (\text{h} | h_m) \propto \text{exp} \Big[ - \tfrac{|\text{h} - \tilde{h}_m^{(i)}|^2}{\tilde{\sigma}_{\tilde{h}:m}^{2(i)}} \Big],
\label{eq:ell_extrinsic_belief_PDA}
\end{equation}
which yields
\begin{subequations}
\label{eq:mean_and_var_extrinsic_belief_PDA}
\begin{equation}
\tilde{h}_m^{(i)} \triangleq \frac{1}{\eta_m^{(i)}} \mathbf{e}_m\herm \mathbf{\Sigma}^{-1(i)} \tilde{\mathbf{y}}_{h:m}^{(i)}\;\;\text{and}\;\;
\tilde{\sigma}_{\tilde{h}:m}^{2(i)} \triangleq \frac{1 - \eta_m^{(i)} \hat{\sigma}_{h:m}^{2(i)}}{\eta_m^{(i)}},
\end{equation}
where $\eta_m^{(i)}$ is a normalization factor defined as
\begin{equation}
\label{eq:eta_PDA}
\eta_m^{(i)} \triangleq \mathbf{e}_m\herm \mathbf{\Sigma}^{-1(i)} \mathbf{e}_m,
\end{equation}
and the common conditional covariance matrix\footnote{The matrix inversion lemma \cite{Ito_ICC_2021} is used in the derivation of equation \eqref{eq:mean_and_var_extrinsic_belief_PDA} and by consequence \eqref{eq:eta_PDA}, such that the same inverse matrix $\mathbf{\Sigma}^{(i)}$ can be used instead of $\mathbf{\Sigma}_m^{(i)}$.}
is given by
\begin{equation}
\mathbf{\Sigma}^{(i)} \triangleq \sum_{m=1}^{K_\tau D_\nu} \hat{\sigma}_{h:m}^{2(i)} \mathbf{e}_m \mathbf{e}_m\herm + N_0 \mathbf{I}_N.
\end{equation}
\end{subequations}

\subsubsection{Soft RG}

Similarly to Section \ref{sec:Generalized_JCDE_at_UE}, except that approximations hold under \ac{VGA} as opposed to \ac{SGA}, the soft replicas of $h_m$ can be inferred from the conditional expectation given the extrinsic beliefs and that the effective noise components in $\hat{h}_m^{(i)}, \forall m$ are uncorrelated, which yields
\begin{equation}
p_{h_{m}\! | \text{h}} ( h_{m} | \text{h}; \bm{\theta}^{(i)})\! = \!\frac{ p_{\text{h} | h_{m}} \!(\text{h} | h_{m};\!\tilde{h}_m^{(i)}\!,\tilde{\sigma}_{\tilde{h}:m}^{2(i)}) \; p_{h_{m}}\!(h_{m}; \bm{\theta}^{(i)}) }{\! \int_{h'_{m}}\! p_{\text{h} | h_{m}} \!(\text{h} | h'_{m};\!\tilde{h}_m^{(i)},\tilde{\sigma}_{\tilde{h}:m}^{2(i)}) \; p_{h_{m}} \!(h'_{m};\! \bm{\theta}^{(i)}) }.
\label{eq:VGA_pdf_h_m_PDA}
\end{equation}

Next, leveraging the assumption that $h_m$ follows a Bernoulli-Gaussian distribution, and using the Gaussian-\ac{PDF} multiplication rule \cite{Parker_TSP_2014}, equation \eqref{eq:VGA_pdf_h_m_PDA} can be rewritten as \cite{VilaTSP2013}
\begin{equation}
p_{h_{m} | \text{h}} ( h_{m} | \text{h}; \bm{\theta}^{(i)})\! = \!(1\! -\! \hat{\rho}_{m}^{(i)}) \delta(h_m)\! +\! \hat{\rho}_{m}^{(i)}\mathcal{CN}(h_m;\hat{h}_{m}^{(i)},\hat{\sigma}_{h:m}^{2(i)}),
\label{eq:VGA_pdf_h_m_rewritten_BG_PDA}
\end{equation}
where
\begin{equation}
\vspace{-1ex}
\label{eq:BG_update_sparsity_rate}
\hat{\rho}_{m}^{(i)} \triangleq \Bigg( \frac{1\! -\! {\rho}^{(i)}}{{\rho}^{(i)}}  \frac{\tilde{\sigma}_{\tilde{h}:m}^{2(i)}\! +\! {\bar{\sigma}}^{(i)}}{\tilde{\sigma}_{\tilde{h}:m}^{2(i)}} \: e^{- \frac{|\tilde{h}_{m}^{(i)}|^2}{\tilde{\sigma}_{\tilde{h}:m}^{2(i)}} + \frac{|\tilde{h}_{m}^{(i)} - {\bar{h}}^{(i)}|^2}{\tilde{\sigma}_{\tilde{h}:m}^{2(i)} + {\bar{\sigma}}^{(i)}}} \!\!+ 1 \Bigg)^{\!\!-1}\!\!\!\!,
\end{equation}
with
\begin{equation}
\label{eq:BG_update_rules}
\hat{h}_{m}^{(i)} \triangleq \frac{{\bar{\sigma}}^{(i)} \tilde{h}_{m}^{(i)} + \tilde{\sigma}_{\tilde{h}:m}^{2(i)} {\bar{h}}^{(i)}}{\tilde{\sigma}_{\tilde{h}:m}^{2(i)} + {\bar{\sigma}}^{(i)}}\;\;\text{and}\;\;
\hat{\sigma}_{h:m}^{2(i)} \triangleq \frac{{\bar{\sigma}}^{(i)} \tilde{\sigma}_{\tilde{h}:m}^{2(i)}}{\tilde{\sigma}_{\tilde{h}:m}^{2(i)} + {\bar{\sigma}}^{(i)}}.
\end{equation}

Finally, adhering to the prior defined for the channel coefficient $h_p$, we choose \ac{wlg}, a denoiser without the mean parameter, namely, ${\bar{h}}^{(i)} = 0$ in equation \eqref{eq:h_m_pdf} and consequently, in equation \eqref{eq:BG_update_rules}\footnote{For completeness, the \ac{EM} parameter update detailed in the consequent section will be derived for all the parameters.}.
From \eqref{eq:VGA_pdf_h_m_rewritten_BG_PDA}, the soft replica $\hat{h}_{m}^{(i)}$ and its \ac{MSE} $\hat{\sigma}_{h:m}^{2(i)}$ can be in general obtained from the conditional expectation as\footnote{Note the already incorporated damping procedure as also done in Section \ref{sec:Generalized_JCDE_at_UE} to prevent convergence to local minima due to incorrect hard-decision replicas.}
\begin{subequations}
\label{eq:soft_rep_and_MSE_updates}
\begin{equation}
\label{eq:PDA_soft_rep_update}
\hat{h}_m^{(i)} = \tilde{\beta}_h \hat{\rho}_{m}^{(i)} \hat{h}_{m}^{(i)} + (1 - \tilde{\beta}_h) \hat{h}_m^{(i-1)},
\end{equation}
\begin{equation}
\label{eq:PDA_MSE_update}
\hat{\sigma}_{h:m}^{2(i)} = \tilde{\beta}_h \big[(1 - \hat{\rho}_{m}^{(i)}) \hat{\rho}_{m}^{(i)} |\hat{h}_{m}^{(i)}|^2 + \hat{\rho}_{m}^{(i)} \hat{\sigma}_{h:m}^{2(i)}\big] + (1-\tilde{\beta}_h) \big[ \hat{\sigma}_{h:m}^{2(i-1)} \big].
\end{equation}
\end{subequations}

\vspace{-1ex}
\subsubsection{Parameter Update via EM}

In order to update the parameter set $\bm{\theta}$ of the Bernoulli-Gaussian distribution at each iteration, the \ac{EM} algorithm is utilized.
In particular, we employ \ac{EM} as an iterative parameter tuning technique to obtain the parameter vector $\bm{\theta}$ that maximizes the likelihood function $p_{\mathbf{y} | \bm{\theta}} (\mathbf{y}|\bm{\theta})$, where $\mathbf{y}$ is defined in equation \eqref{eq:modified_input_output_relation_for_estimation}, with $\bm{\theta}$ as given in equation \eqref{eq:h_m_pdf}.

To this end, we follow \cite{BishopBook2006} and first convert $p_{\mathbf{y} | \bm{\theta}} (\mathbf{y}|\bm{\theta})$ to a tractable form by utilizing latent variables, marginalizing over an arbitrary \ac{PDF} $\hat{p}_{\mathbf{{h}}} (\mathbf{h})$ and using the log-likelihood function $\text{ln} \big[ p_{\mathbf{y} | \bm{\theta}} (\mathbf{y}|\bm{\theta})\big]$, reformulated to highlight the differential entropy and \ac{KL} divergence terms $\psi(\mathbf{h})$ and $\mathrm{D}_\text{KL}\big(\hat{p}_{\mathbf{{h}}} (\mathbf{h}) || p_{\mathbf{h} | \mathbf{y} , \bm{\theta}} (\mathbf{h} | \mathbf{y},\bm{\theta})\big)$, respectively, namely
\begin{eqnarray}
\text{ln} \big[ p_{\mathbf{y} | \bm{\theta}} (\mathbf{y}|\bm{\theta})\big] \propto \int_\mathbf{h} \Big[ \hat{p}_{\mathbf{{h}}} (\mathbf{h}) \, \text{ln} \, p_{\mathbf{{h}},\mathbf{y} , \bm{\theta}} (\mathbf{h},\mathbf{y},\bm{\theta}) + \psi(\mathbf{h}) &&
\nonumber \\
&& \hspace{-33ex}  + \mathrm{D}_\text{KL}\big(\hat{p}_{\mathbf{{h}}} (\mathbf{h}) || p_{\mathbf{h} | \mathbf{y} , \bm{\theta}} (\mathbf{h} | \mathbf{y},\bm{\theta})\big) \Big].
\label{eq:log_likelihood_function_for_I}
\end{eqnarray}

Next, let us define the auxiliary function
\begin{equation}
\mathrm{J} \big( \hat{p}_{\mathbf{{h}}} (\mathbf{h}), \bm{\theta} \big) \triangleq \int_\mathbf{h} \Big[ \hat{p}_{\mathbf{{h}}} (\mathbf{h}) \, \text{ln} \, p_{\mathbf{{h}},\mathbf{y} , \bm{\theta}} (\mathbf{h},\mathbf{y},\bm{\theta}) + \psi(\mathbf{h}) \Big].
\label{eq:def_J_N_div}
\end{equation}

Then, taking advantage of the non-negativity of the \ac{KL} divergence, the following lower bound on the latter log-likelihood function, can be obtained as
\begin{equation}
\text{ln} \big[ p_{\mathbf{y} | \bm{\theta}} (\mathbf{y}|\bm{\theta})\big] \geq \mathrm{J} \big( \hat{p}_{\mathbf{{h}}} (\mathbf{h}), \bm{\theta} \big).
\label{eq:non_neg_N_prop}
\end{equation}

The \ac{EM} algorithm summarized above allows us to maximize the log-likelihood function in a cost-effective manner by alternating between the E-step minimizing \ac{KL} divergence in equation \eqref{eq:log_likelihood_function_for_I} and the M-step maximizing the lower bound of the log-likelihood in equation \eqref{eq:non_neg_N_prop}, given by
\begin{subequations}
\begin{eqnarray}
\label{eq:E-step}
&&\hspace{-5ex}\text{E-step:}\quad \hat{p}_{\mathbf{{h}}} (\mathbf{h}) = \underset{\hat{p}'_{\mathbf{{h}}} (\mathbf{h})}{\mathrm{arg \; min}} \; J \big( \hat{p}'_{\mathbf{{h}}} (\mathbf{h}), \bm{\theta} \big),\\
\label{eq:M-step}
&&\hspace{-5ex}\text{M-step:}\quad \bm{\theta} = \underset{\bm{\theta}'}{\mathrm{arg \; max}} \; J \big( \hat{p}_{\mathbf{{h}}} (\mathbf{h}), \bm{\theta}' \big).
\end{eqnarray}
\end{subequations}

Regarding the E-step, notice that solving equation \eqref{eq:E-step} is equivalent to minimizing of $\hat{p}_{\mathbf{{h}}} (\mathbf{h})$ given a fixed $\bm{\theta}$, but equation \eqref{eq:VGA_pdf_h_m_rewritten_BG_PDA} already computes the solution to this problem.
In other words, the Bernoulli-Gaussian distribution in  equation \eqref{eq:VGA_pdf_h_m_rewritten_BG_PDA} is the closed-form solution of problem \eqref{eq:E-step}.

In turn, the focus of the M-step is to maximize $\bm{\theta}$ for a given distribution $\hat{p}_{\mathbf{{h}}} (\mathbf{h})$. Using \eqref{eq:VGA_pdf_h_m_rewritten_BG_PDA}, the maximization problem in \eqref{eq:M-step} can be reformulated as
\begin{equation}
\bm{\theta} = \underset{\bm{\theta}'}{\mathrm{arg \; max}} \; \mathbb{E}_{\mathbf{h} | \text{h}} \Big\{ \text{ln} \big[ p_{\mathbf{{h}},\mathbf{y} , \bm{\theta}} (\mathbf{h},\mathbf{y},\bm{\theta}') \big] |\; \text{h} ; \bm{\theta}^{(i)}  \Big\},
\label{eq:reform_M-step}
\end{equation}
and solved efficiently via a Lagrange method, yielding the update rules
\begin{subequations}
\begin{equation}
{\rho}^{(i)} = \frac{1}{K_\tau D_\nu} \sum_{m=1}^{K_\tau D_\nu} \hat{\rho}_{m}^{(i)},
\label{eq:Update_for_lambda}
\vspace{-0.75ex}
\end{equation}
\begin{equation}
{\bar{h}}^{(i)} = \frac{1}{{K_\tau D_\nu} \cdot {\rho}^{(i)}} \sum_{m=1}^{K_\tau D_\nu} \hat{\rho}_{m}^{(i)} \hat{h}_{m}^{(i)},
\label{eq:Update_for_mu}
\vspace{-0.75ex}
\end{equation}
\begin{equation}
{\bar{\sigma}}^{(i)} = \frac{1}{{K_\tau D_\nu} \cdot {\rho}^{(i)}} \sum_{m=1}^{K_\tau D_\nu} \hat{\rho}_{m}^{(i)} \bigg(  \big|  \hat{h}_{m}^{(i)} - {\bar{h}}^{(i)}  \big|^2 + \hat{\sigma}_{h:m}^{2(i)} \bigg).
\label{eq:Update_for_phi}
\vspace{-0.5ex}
\end{equation}
\label{eq:Complete_set_update_BG}
\end{subequations}

\vspace{-1ex}
\begin{algorithm}[t!]
\caption{PDA-based Radar Parameter Estimation (Prop.)}
\label{alg:PDA-EM}
\textbf{Input:} Receive signal $\mathbf{y}$, dictionary matrix $\mathbf{E}$, noise power $N_0$, number of paths $P$, average channel power per path $\sigma_h^2$, maximum number of iterations $i_\text{max}$ and damping factor $\tilde{\beta}_h$. \\
\textbf{Output:} Estimates $\hat{\tau}_p$ and $\hat{\nu}_p$ extracted from the non-zero indices of the sparse channel estimate vector $\!\hat{\;\mathbf{h}}$.\\
\textbf{Initialization:} Set counter to $i = 0$, set initial distribution parameters ${\rho}^{(0)} = P/(K_\tau D_\nu)$ and ${\bar{\sigma}}^{(0)} = 1/P$, set average channel power per path $\sigma_h^2 = 1/(K_\tau D_\nu)$, and set initial estimates $\hat{h}_{m}^{(0)} = 0$ and ${\hat{\sigma}}_{h:m}^{(0)} = \sigma_h^2, \forall m$.
\vspace{0.4ex}
\hrule

\begin{algorithmic}[1]

\STATEx \hspace{-3.8ex}{\textbf{for}} {$i=1$ to $i_\text{max}$} {\textbf{do}} $\forall m$
\STATE Compute soft signal vectors $\tilde{\mathbf{y}}_{h:m}^{(i)}$ from equation \eqref{eq:Soft_IC_PDA}.
\STATE Compute soft extrinsic channel beliefs $\tilde{h}_m^{(i)}$ and their
variances $\tilde{\sigma}_{\tilde{h}:m}^{2(i)}$ from equation \eqref{eq:mean_and_var_extrinsic_belief_PDA}.
\STATE Compute denoised sparsity rates $\hat{\rho}_m^{(i)}$ from eq. \eqref{eq:BG_update_sparsity_rate}.
\STATE Compute denoised channel estimates $\hat{h}_{m}^{(i)}$  and their variances $\hat{\sigma}_{h:m}^{2(i)}$ from equation \eqref{eq:BG_update_rules}.
\STATE Compute damped channel estimates $\hat{h}_{m}^{(i)}$ and variances $\hat{\sigma}_{h:m}^{2(i)}$ from equation \eqref{eq:soft_rep_and_MSE_updates}.
\STATE Update distribution parameters ${\rho}^{(i)}$ and ${\bar{\sigma}}^{(i)}$ from eq. \eqref{eq:Complete_set_update_BG}.
\STATEx \hspace{-3.8ex}{\textbf{end}} {\textbf{for}}
\end{algorithmic}
\label{PDA-EM}
\noindent Compute the estimates $\hat{\tau}_p$ and $\hat{\nu}_p$ corresponding to the indices $m$ of the non-zero entries of $\!\hat{\;\mathbf{h}}$ in accordance to equation \eqref{eq:sparse_channel_vector}.
\end{algorithm}

A complete and compact description of the proposed \ac{PDA}-based \ac{RPE} scheme is given in the form of a pseudocode in Algorithm \eqref{alg:PDA-EM}.
Upon convergence of the above procedure, the estimate of the sparse vector $\mathbf{h}$ is known. Finally, the $(i,j)$ indices of $\mathbf{h}$ can be used to obtain the corresponding delay and Doppler shifts, and hence the corresponding ranges and velocities of the objects in the vicinity.

\noindent \textit{Remark 2:} Due to the nature of the \ac{RPE} formulation, a state evolution analysis \cite{Donoho2009,LiuTIT2023} is no longer possible since the equivalent observation matrix is not \ac{i.i.d.} sub-Gaussian distributed and/or right-unitary invariant with a finite system size.
In addition, in order to handle the quite strong correlation amongst the elements of the equivalent observation matrix, the proposed \ac{RPE} algorithm is designed based on the \ac{PDA} framework \cite{TakahashiTWC2022}, which while robust against such observation correlations, does not guarantee Bayesian optimality in the large-system limit. 
%
If the algorithm was designed via a Bayes-optimal technique such as \ac{GAMP} \cite{RanganISITP2011} or \ac{EP} \cite{TakeuchiTIT2020}, the Onsager correction term will behave in an unstable manner and the estimation performance will deteriorate significantly. 
Given the above, it is difficult to perform a theoretical convergence analysis; hence, we have verified the number of iterations required for convergence via a thorough numerical analysis as shown below (Figure \ref{fig:convergence_plot}).

\vspace{-2ex}
\subsection{Complexity Analysis}
\label{sec:RPE_Complexity}

The computational complexity for both the proposed \ac{PDA}-based \ac{RPE} scheme detailed in Algorithm \ref{alg:PDA-EM}, and the \ac{SotA} \ac{SBL}-based \ac{RPE} scheme given in Algorithm \ref{alg:SBL-EM}, are dominated by the matrix inversion on line 2 of both algorithms, required at every iteration of the corresponding algorithms, albeit of different sizes in each. 

While the \ac{SotA} method has a computational complexity of order $\mathcal{O}((K_\tau D_\nu)^3)$, the complexity order of the proposed method is $\mathcal{O}(N^3)$.
In most cases, since $N \approx K_\tau \cdot D_\nu$, the overall computational complexity is very similar.
It is also noteworthy that this translates to the proposed \ac{PDA}-based \ac{RPE} technique being almost completely independent of the grid size based on the values of $K_\tau$ and $D_\nu$, which means that much larger grids (even over multiple iterations as in \cite{Ranasinghe_ICASSP_2024}) for high accuracy can be used with the proposed method as opposed to the \ac{SotA} which incurs a heavy computational burden due to the matrix inversion dependent on $K_\tau \cdot D_\nu$.

Finally, we emphasize that in addition to the performance gain and the similar complexity of the proposed method, we also show that the proposed method converges much faster, as shown in Figure  \ref{fig:convergence_plot}.

\vspace{-2ex}
\subsection{Performance Assessment}

\addtocounter{footnote}{-1}
In order to evaluate the performance of the proposed \ac{PDA}-based \ac{RPE} algorithm, we consider the monostatic scenario depicted in Figure \ref{fig:ISAC_system_model}. 
Under such a scenario, and in light of the results of Figure \ref{fig:XXXM_Performance_plot} -- as well as related literature, $e.g.$\cite{GaudioTWC2020}

\vspace{-1ex}
\begin{figure}[H]
\centering
\captionsetup[subfloat]{labelfont=small,textfont=small}
\subfloat[Range Estimation Performance (\ac{FD} system, single target).]{{\includegraphics[width=\columnwidth]{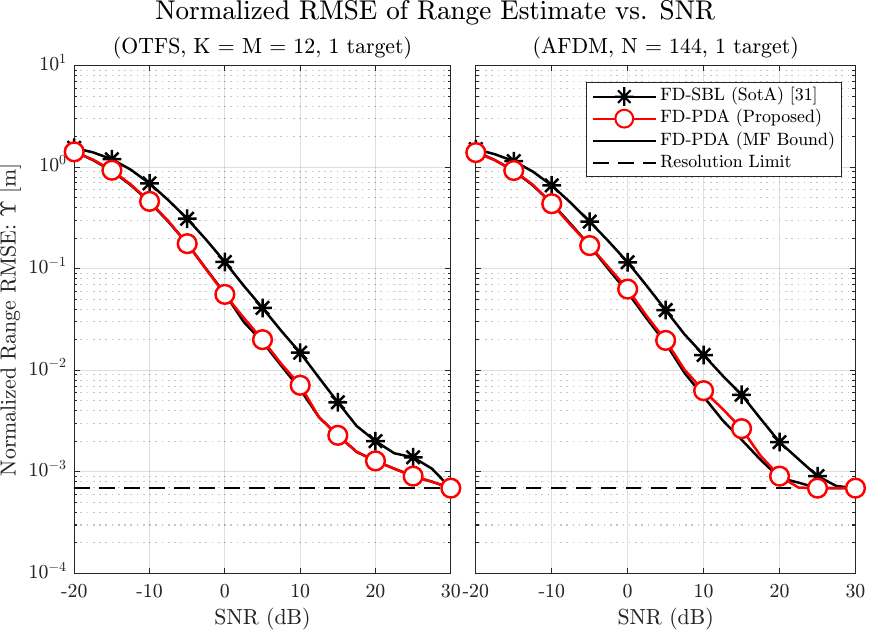}}}%
\label{fig:OTFS_vs_AFDM_Range}
\subfloat[Velocity Estimation Performance (\ac{FD} system, single target).]{{\includegraphics[width=\columnwidth]{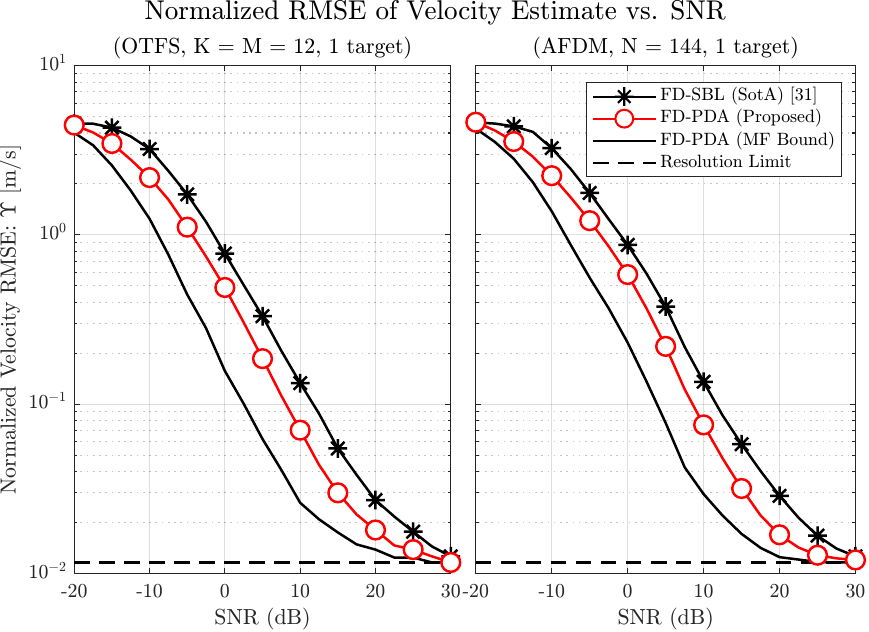}}}%
\label{fig:OTFS_vs_AFDM_Velocity}
\vspace{-2ex}
\caption[]{\ac{RMSE} versus \ac{SNR}\footnotemark performance of \ac{SotA} \ac{SBL} and proposed \ac{PDA} monostatic \ac{RPE} schemes over \ac{OTFS} and \ac{AFDM} waveforms in a \ac{FD} (interference-free) scenario with a single \ac{LoS} signal from a single target.}
\label{fig:OTFS_vs_AFDM}
\end{figure}
\vspace{-1ex}
 
\noindent  -- which demonstrate that \ac{OFDM} is outperformed by \ac{OTFS} and \ac{AFDM} in doubly-dispersive channels, we consider hereafter only \ac{OTFS} and \ac{AFDM} systems.

Our first set of results, shown in Figure \ref{fig:OTFS_vs_AFDM}, are obtained for a single \ac{LoS} reflected signal, from a single target, positioned at $15$ m and traveling at a velocity of $151$ km/h towards the \ac{BS}, with the remark that since both the \ac{SotA} \ac{SBL} and the proposed \ac{PDA} \ac{RPE} algorithms are based on the sparse model described by the set of equations \eqref{eq:sparse_RPE}, the estimation algorithm themselves have no hard restrictions on the number of paths or targets to be estimated, as long as the grid modeled by the dictionary matrix $\mathbf{E}$ in equation \eqref{eq:sparse_dict_matrix_def} is fine enough to ensure a sufficient sparsity of the vector $\mathbf{y}$ in equation \eqref{eq:modified_input_output_relation_for_estimation}.

With that clarified, the motivation of assessing the performance of these algorithms under a single path/target case is to grant an advantage to the \ac{OTFS} approach, where a larger number of targets only increases the weight of the assumption of no self-interference (not required in the \ac{AFDM} case).

\begin{figure}[t!]
  \centering
  {%
  \includegraphics[width=\columnwidth]{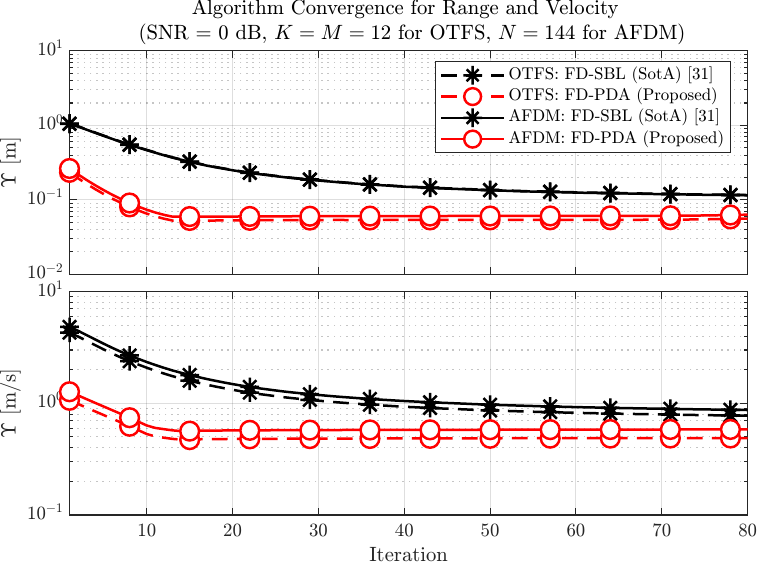}}
  \vspace{-4ex}
  \caption{Convergence of \ac{SotA} \ac{SBL} and proposed \ac{PDA} monostatic \ac{RPE} schemes over \ac{OTFS} and \ac{AFDM}.}
  \label{fig:convergence_plot}
  \captionsetup[subfloat]{labelfont=small,textfont=small}
  \subfloat[Range Estimation Performance (\ac{SIC}-enabled system, multiple targets).]{{\includegraphics[width=\columnwidth]{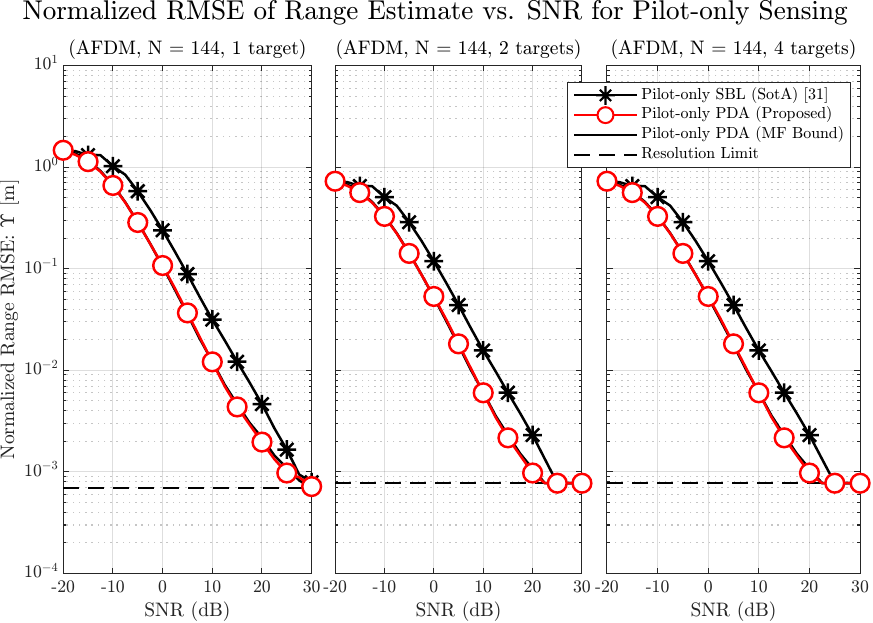}}}%
  \label{fig:SIC_range_plot}
  \subfloat[Velocity Estimation Performance (\ac{SIC}-enabled system, multiple targets).]{{\includegraphics[width=\columnwidth]{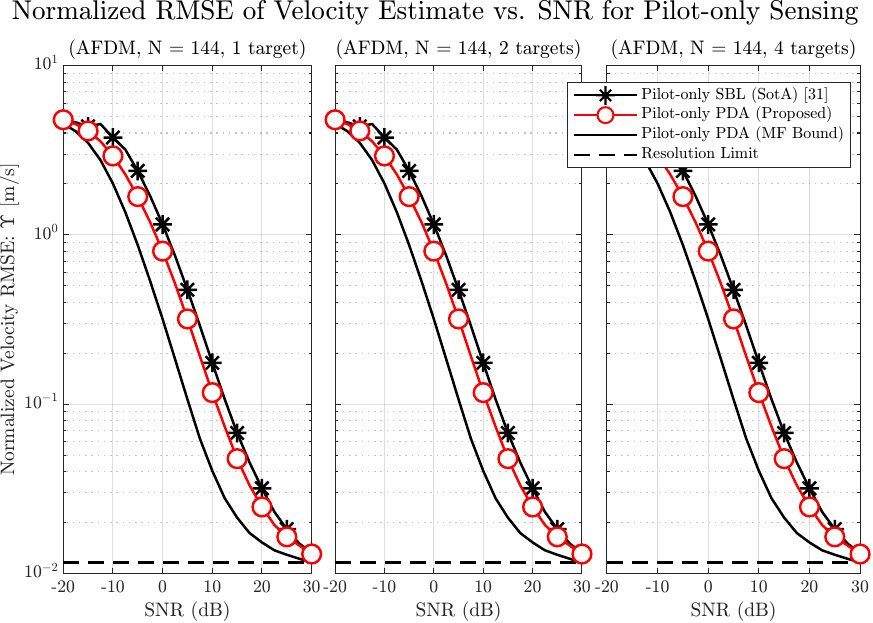}}}%
  \label{fig:SIC_vel_plot}
  \vspace{-1ex}
  \caption[]{\ac{RMSE} versus \ac{SNR} performance of \ac{SotA} \ac{SBL} and proposed \ac{PDA} monostatic \ac{RPE} schemes over \ac{AFDM} waveform, executed using only the single pilot and null-guard interval enabling \ac{SIC} via a simple low-complexity analog dechirping and filtering\cite{Bemani_WCL_2024}, in a scenario with multiple targets.}
  \label{fig:SIC_radar_param_plots}
  \vspace{-2ex}
\end{figure}

The remaining system parameters for the results shown in Figure \ref{fig:OTFS_vs_AFDM} are in line with those used when assessing \ac{JCDE} schemes in Subsection \ref{sec:JCDE_Simulations}, namely, $70 \, \text{GHz}$ band, with bandwidth of $20 \, \text{MHz}$, \ac{QPSK} modulation, and maximum normalized delay and digital Doppler shift indices $\ell_\text{max} = 20$ and $f_\text{max} = 0.25$, respectively.
In order to reduce complexity, we reduce the system size to $K=B = M = 12$ for \ac{OTFS}, and $N = 144$ for \ac{AFDM}, which results in an unambiguous maximum range of $75$ m and velocity of $535$ km/h.
Finally, the proposed \ac{PDA}-based \ac{RPE} algorithm employs a damping factor $\tilde{\beta}_h = 0.5$ and is run up to $i_\text{max} = 40$ iterations, while the \ac{SBL} scheme is run up to $i_\text{max} = 80$ iterations, due to its slower convergence (see Figure \ref{fig:convergence_plot}).

\footnotetext{While many formulations for the \ac{SNR} exist in literature, including radar-centric metrics as used in \cite{GaudioTWC2020, Ranasinghe_ICASSP_2024}, we use the typical definition hinging on $h_p$ as used in \cite{Bemani_WCL_2024}, which already incorporates the effect of the radar cross-section parameter $\sigma_\text{rcs} = 1$ denoting that no power is lost due to the reflection.}

The performance metric adopted is the normalized \ac{RMSE}, averaged over all targets and realizations, which is defined as
\vspace{-1.5ex}
\begin{equation}
\Upsilon \triangleq \mathbb{E}\Big[\frac{1}{P+1}\sum_{p=0}^P\frac{||\hat{\vartheta}_p - \vartheta_p||^2_2}{|\vartheta_p|}\Big],
\label{eq:Normalized_RMSE}
\vspace{-0.5ex}
\end{equation}
where $\vartheta_p$ denotes a given radar parameter (range or velocity), $\hat{\vartheta}_p$ its estimate, and the expectation is taken over the number of realizations of the algorithm.

Finally, Figure \ref{fig:OTFS_vs_AFDM} also includes plots of the resolution limit\footnote{In this case, the resolution limit defines how fine the search grid is; $i.e.,$ it measures the estimation accuracy for the range and velocity if the estimate for $\!\hat{\;\mathbf{h}}$, obtained as the output of the \ac{SotA} and proposed algorithms, is perfect.} \cite{Ranasinghe_ICASSP_2024} and the \ac{MF} bound \cite{XiaochenOJCS2023}, obtained by executing Algorithm \ref{alg:PDA-EM} for a single iteration, initialized with the true radar parameter values.
The results clearly indicate that the proposed \ac{PDA} method outperforms the \ac{SotA} \ac{SBL} approach both for range and velocity estimation, both under \ac{OTFS} and \ac{AFDM} waveforms, with the proposed method reaching the \ac{MF} bound in the case of range estimates.
It is also found, moreover, that \ac{AFDM} yields slightly better performance than \ac{OTFS}, and that all algorithms reach the resolution bound under a sufficiently large \ac{SNR}, indicating that both methods converge absolutely.

In order to clarify that convergence is also not an issue at low \ac{SNR}, we offer in Figure \ref{fig:convergence_plot} plots of the range and velocity normalized \ac{RMSE}.
It can be seen from those results that the proposed \ac{PDA}-based method is advantageous compared to the \ac{SotA} \ac{SBL} approach also in terms of convergence speed.

Having established the overall advantage of \ac{AFDM} over \ac{OTFS} in \ac{FD} scenarios, and of the proposed \ac{PDA}-based \ac{RPE} algorithm over the \ac{SBL}-based \ac{SotA} alternative, we finally confront the two critical issues left unaddressed in the performance assessment so far, namely, the impact of self-interference and a larger number of targets to be estimated.
To that end, we offer in Figure \ref{fig:SIC_radar_param_plots} plots equivalent to those shown in Figure \ref{fig:OTFS_vs_AFDM}, but this time exclusively for \ac{SIC}-enabled \ac{AFDM} systems, and in scenarios with multiple targets using only the single pilot and null-guard interval for \ac{RPE}.
Notice that this only requires minimal modification to equation \eqref{eq:sparse_RPE} such that it becomes $\mathbf{y}_p = \mathbf{E}_p \cdot \mathbf{h} + \tilde{\mathbf{w}_p} \in \mathbb{C}^{B \times 1}$ with $\mathbf{y}_p \in \mathbb{C}^{B \times 1}$ and $\mathbf{E}_p \in \mathbb{C}^{B \times K_\tau D_\nu}$ being the parts of the received signal $\mathbf{y} \in \mathbb{C}^{N \times 1}$ and dictionary matrix $\mathbf{E} \in \mathbb{C}^{N \times K_\tau D_\nu}$, respectively, corresponding to the single pilot and the null-guard interval $\mathbf{x}_p \in \mathbb{C}^{B \times 1}$ of the original transmit symbol vector $\mathbf{x} \in \mathbb{C}^{N \times 1}$.
Consequently, both the \ac{SotA} and proposed algorithms apply to monostatic \ac{RPE}, such that the inherent self-interference from the \ac{LoS} path can be cancelled via a simple low-complexity analog dechirping and filtering\cite{Bemani_WCL_2024} as opposed to the costly solutions required for \ac{OFDM} and \ac{OTFS} systems.
\vspace{-2.5ex}
\section{Conclusion}
We offered two new \ac{ISAC}-enabling contributions for the emerging \ac{AFDM} waveform, namely, a \ac{PBiGaBP}-based \ac{JCDE} algorithm, which enables channel estimation with low pilot overhead, and a \ac{PDA}-based \ac{RPE} algorithm, which enables the \ac{BS} to act as a monostatic radar, estimating the delay and Doppler parameters of multiple paths simultaneously. 
Using a generalized channel model under which \ac{OFDM}, \ac{OTFS} and \ac{AFDM} can be directly compared, it was shown that the proposed algorithms outperform \ac{SotA} alternatives and that \ac{AFDM} is advantageous over \ac{OTFS} and \ac{OFDM} both for communications and sensing functionalities in doubly-dispersive channels.

\vspace{-2ex}

  \newpage
  \tiny{.}
  \newpage
  \section*{Appendices/Illustrations}{
    \label{app:JCDE_diagram}
    
    \begin{figure}[H]
    \centering
    \includegraphics[width=1.8\columnwidth]{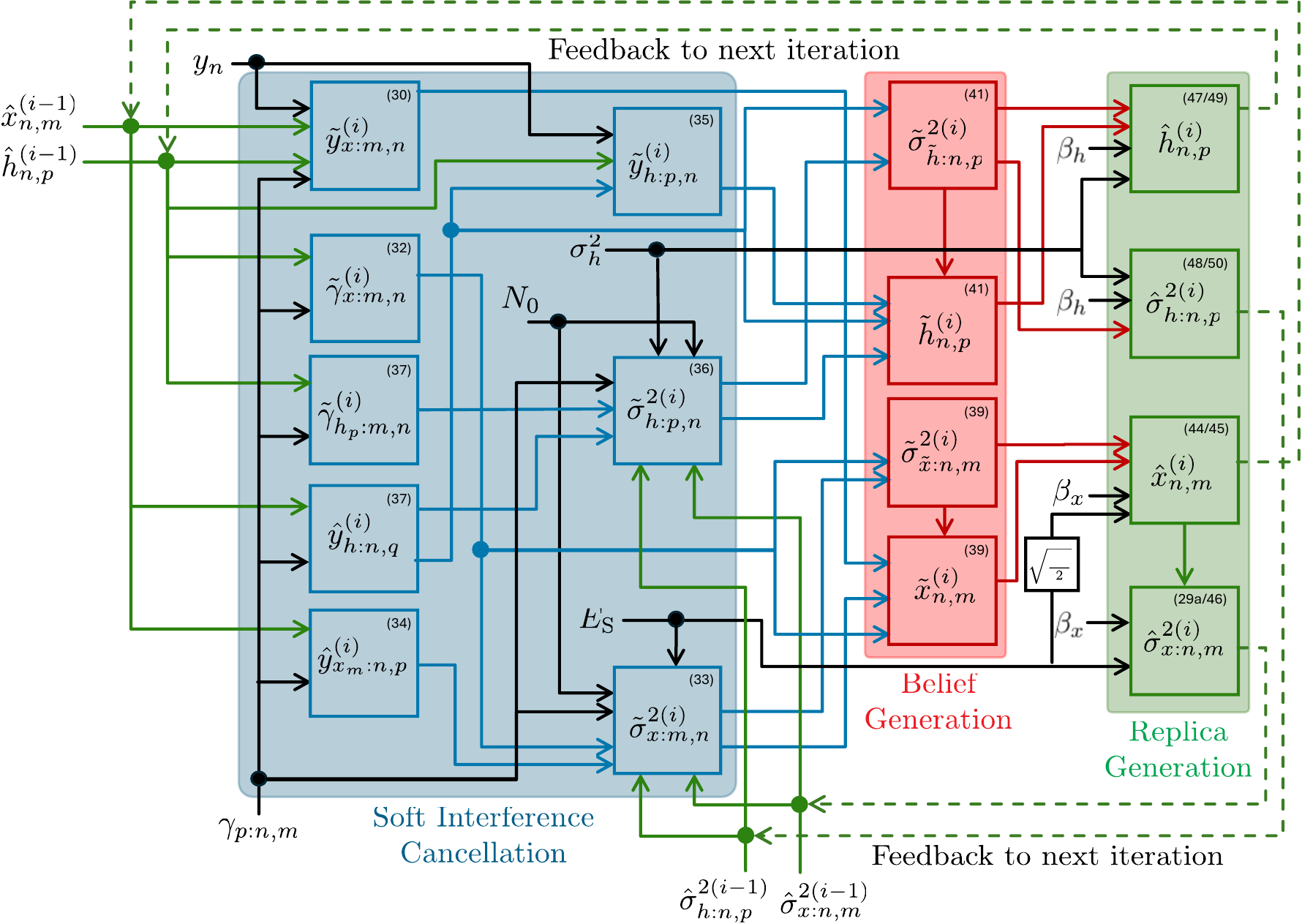}%
    \caption{Diagram of the Proposed \ac{PBiGaBP}-based \ac{JCDE} in Algorithm \ref{alg:JCDE_PBiGaBP}.}
    \label{fig:JCDEDiagram}
    \vspace{-2ex}
    \end{figure}

    \label{app:RPE_diagram}
    
    \begin{figure}[H]
    \centering
    \includegraphics[width=1.8\columnwidth]{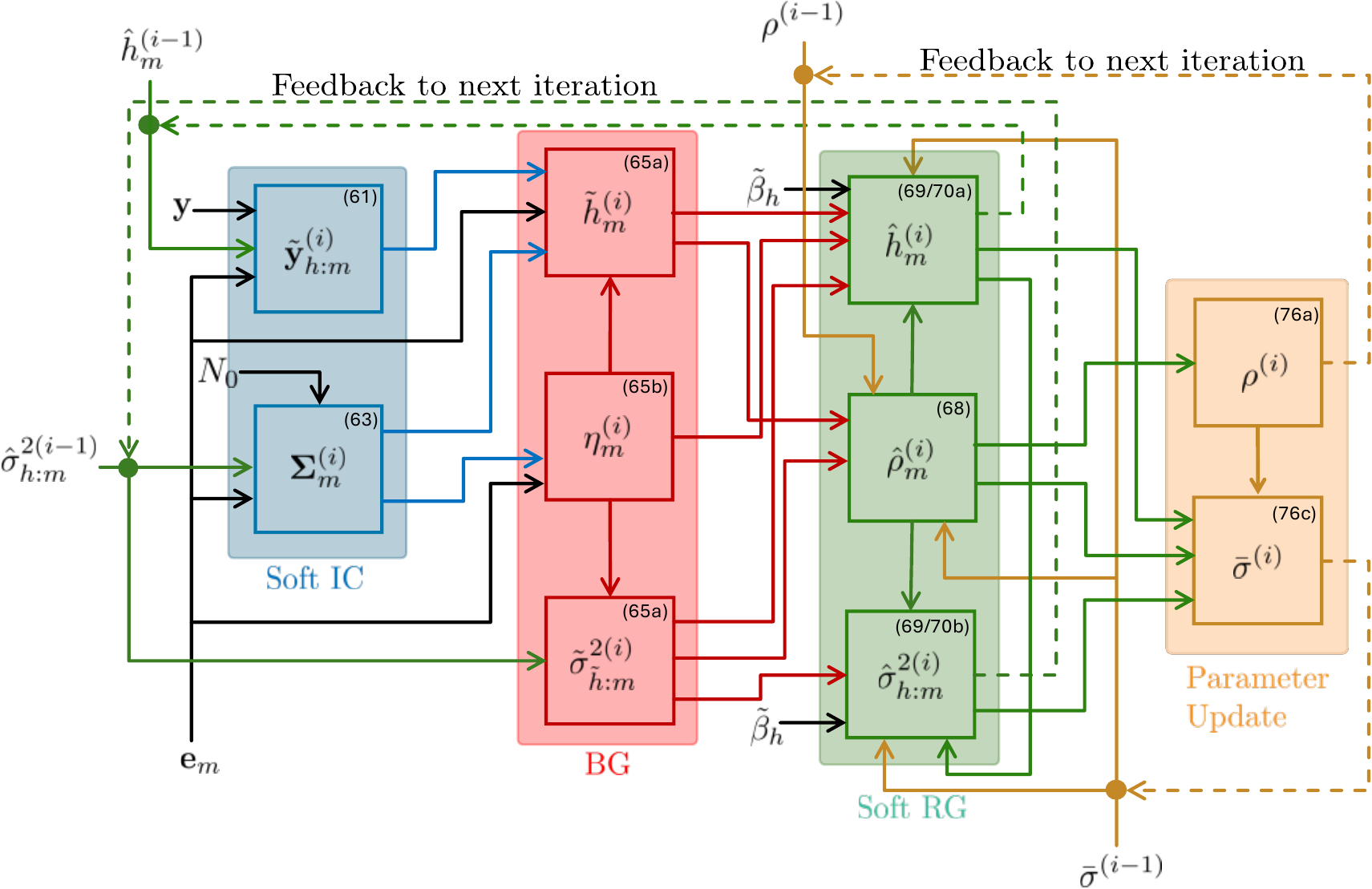}%
    \label{fig:RPEDiagram}
    \caption{Diagram of the Proposed \ac{PDA}-based \ac{RPE} in Algorithm \ref{alg:PDA-EM}.}
    \vspace{-2ex}
    \end{figure}
    }

\cleardoublepage
\begin{IEEEbiography}[{\includegraphics[width=1in,height=1.25in,clip,keepaspectratio]{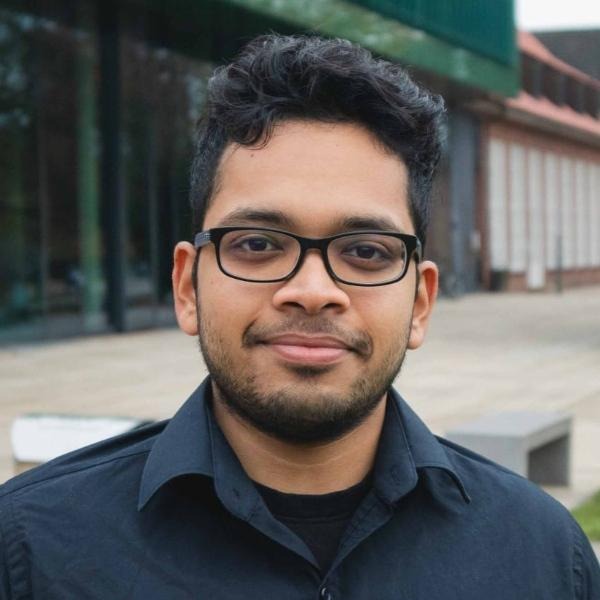}}]{Kuranage Roche Rayan Ranasinghe} (Graduate Student Member, IEEE)
obtained B.Sc. degrees in Electrical and Computer Engineering, as well as Robotics and Intelligent Systems (with a minor in Physics) from Constructor University (formerly Jacobs University Bremen) in 2023/2024, where he is currently pursuing a Ph.D. in Electrical Engineering. Within the fields of wireless communications and signal processing, his research interests encompass integrated sensing, communications and computing (ISCC), compressed sensing, Bayesian inference, belief propagation and optimization theory.
\end{IEEEbiography}
\begin{IEEEbiography}[{\includegraphics[width=1in,height=1.25in,clip,keepaspectratio]{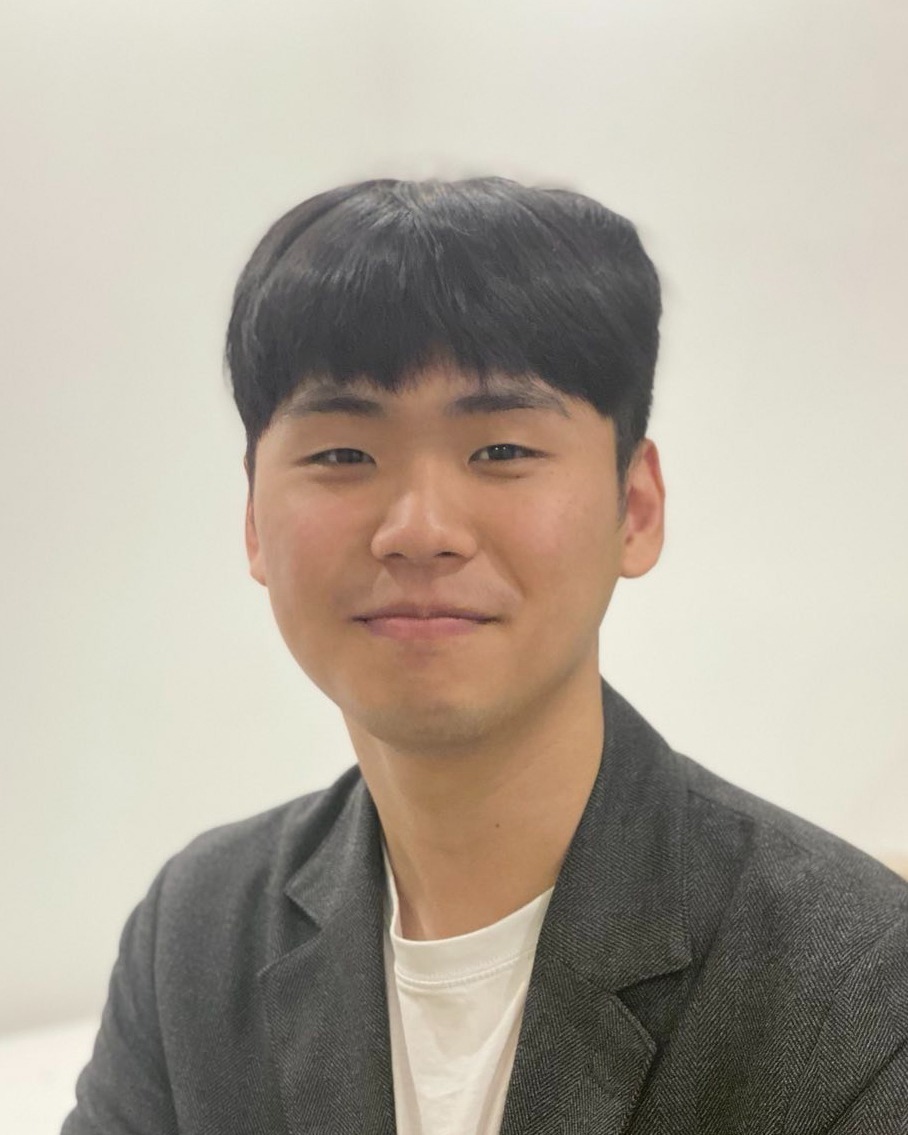}}]{Hyeon Seok Rou} (Graduate Student Member, IEEE)
  has received the Ph.D. degree in Electrical Engineering in 2024 from Constructor University, Bremen, Germany, and the B.Sc. degree in Electrical and Computer Engineering (ECE) in 2021 from Jacobs University Bremen, Germany. He has received the Korea Institute of Science and Technology Europe Research Scholarship Award from The Korean Scientists and Engineers Association in the FRG (Verein Koreanischer Naturwissenschaftler und Ingenieure in der BRD e.V.) in 2022, and he was a visiting researcher at the Intelligent Communications Lab, Korea Advanced Institute of Science and Technology (KAIST) in 2023. His research interests include integrated sensing and communications (ISAC), signal processing in doubly-dispersive channels, high-mobility communication systems, multi-dimensional modulation, next-generation metasurfaces, B5G/6G V2X communication technologies, and quantum computing.
\end{IEEEbiography}

\begin{IEEEbiography}[{\includegraphics[width=1in,height=1.25in,clip,keepaspectratio]{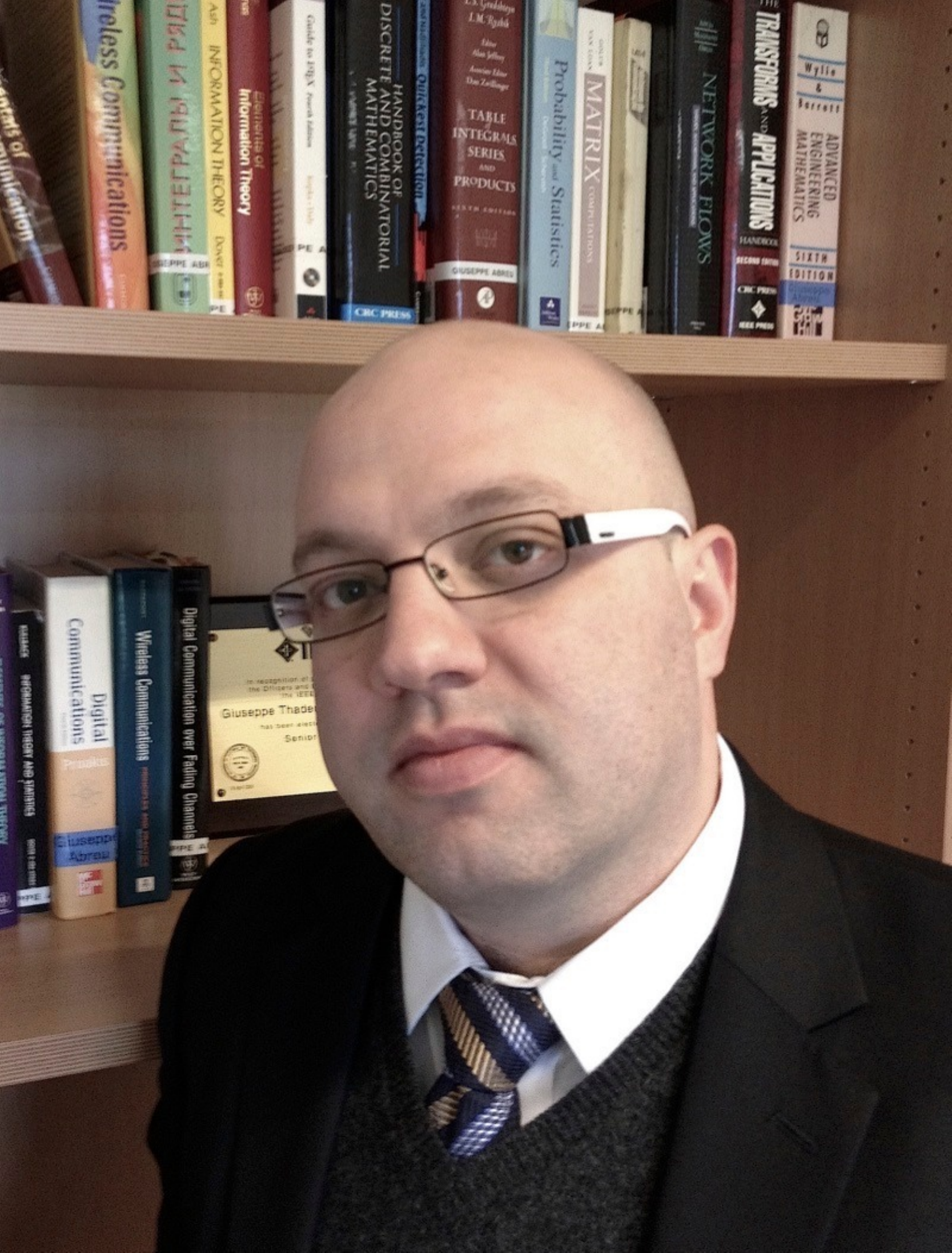}}]{Giuseppe Thadeu Freitas de Abreu} (Senior Member, IEEE) received the B.Eng. degree in electrical engineering and the specialization Latu Sensu
degree in telecommunications engineering from the
Universidade Federal da Bahia (UFBa), Salvador,
Bahia, Brazil in 1996 and 1997, respectively, and
the M.Eng. and D.Eng. degrees in physics, electrical,
and computer engineering from Yokohama National
University, Japan, in March 2001 and March 2004,
respectively. He was a postdoctoral fellow and later
an adjunct professor (docent) in statistical signal
processing and communications theory at the Department of Electrical and
Information Engineering, University of Oulu, Finland from 2004 to 2006
and from 2006 to 2011, respectively. Since 2011, he has been a professor
of electrical engineering at Jacobs University, Bremen, Germany. From
April 2015 to August 2018, he simultaneously held a full professorship
at the Department of Computer and Electrical Engineering, Ritsumeikan
University, Japan. His research interests include communications and signal
processing, including communications theory, estimation theory, statistical
modeling, wireless localization, cognitive radio, wireless security, MIMO
systems, ultrawideband and millimeter wave communications, full-duplex and
cognitive radio, compressive sensing, energy harvesting networks, random
networks, connected vehicles networks, and many other topics. He received
the Uenohara Award at Tokyo University in 2000 for his master’s thesis.
He has been a co-recipient of the Best Paper Award at several international
conferences. He was awarded prestigious JSPS, Heiwa Nakajima, and NICT
Fellowships in 2010, 2013, and 2015, respectively. He served as an associate
editor for the IEEE Transactions on Wireless Communications from 2009 to
2014 and the IEEE Transactions on Communications from 2014 to 2017; and
as an executive editor for IEEE Transactions on Wireless Communications
from 2017 to 2021. He is currently serving as an editor to the IEEE Signal
Processing Letters and the IEEE Communications Letters.
\end{IEEEbiography}

\begin{IEEEbiography}[{\includegraphics[width=1in,height=1.25in,clip,keepaspectratio]{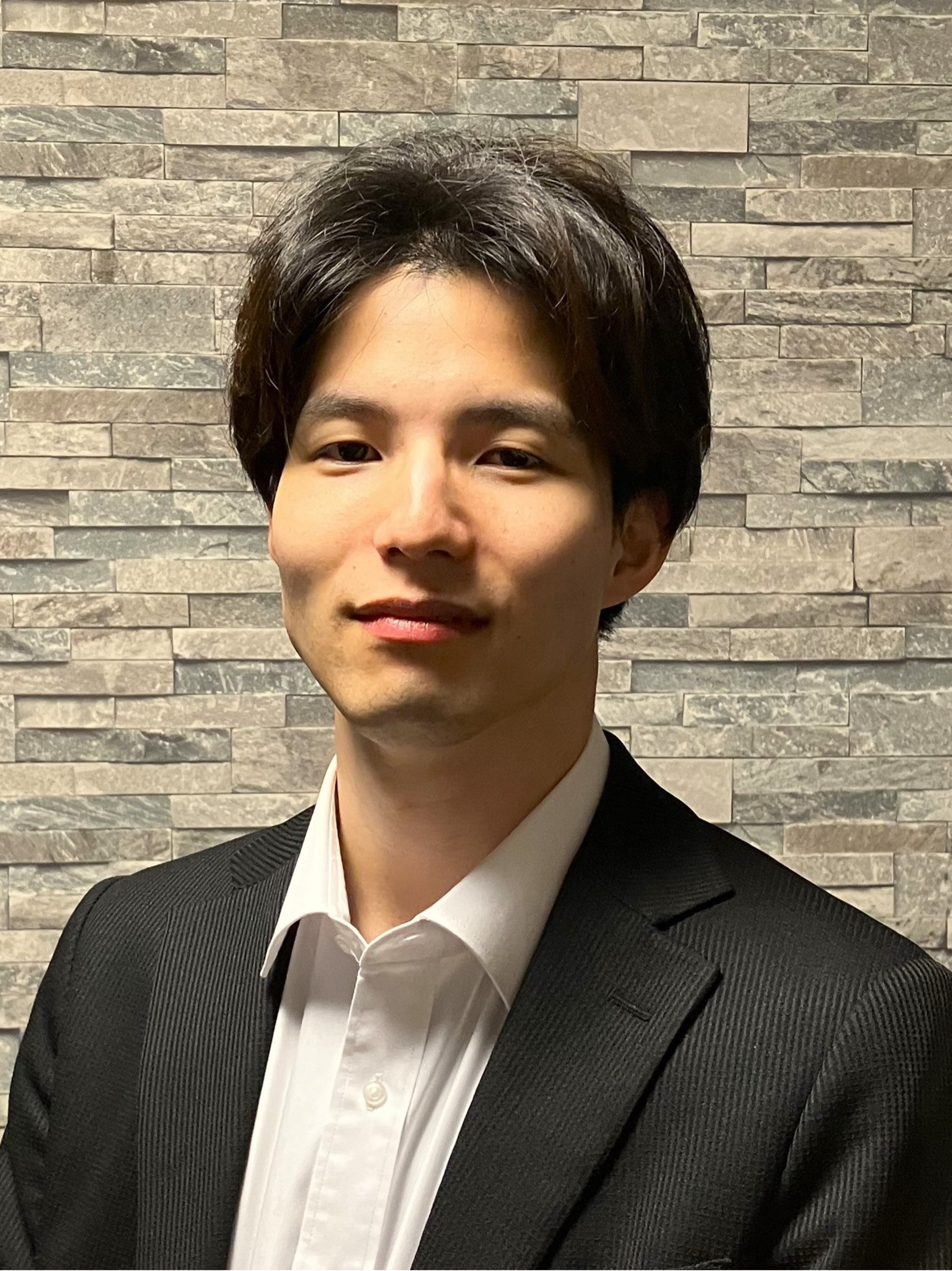}}]{\noindent Takumi Takahashi} (Member, IEEE) received the B.E., M.E., and Ph.D. degrees in communication engineering from Osaka University, Osaka, Japan, in 2016, 2017, and 2019, respectively. From 2018 to 2019, he was a Visiting Researcher at the Centre for Wireless Communications, University of Oulu, Finland. In 2019, he joined the Graduate School of Engineering, Osaka University, as an Assistant Professor. His current research interests include Bayesian inference, belief propagation, signal processing, and wireless communications. He received the 80th Best Paper Award from IEICE and the 2019 and 2023 Best Paper Awards from IEICE Communication Society. He was certified as an Exemplary Reviewer of IEEE Wireless Communications Letters in 2023.
\end{IEEEbiography}

\begin{IEEEbiography}[{\includegraphics[width=1in,height=1.25in,clip,keepaspectratio]{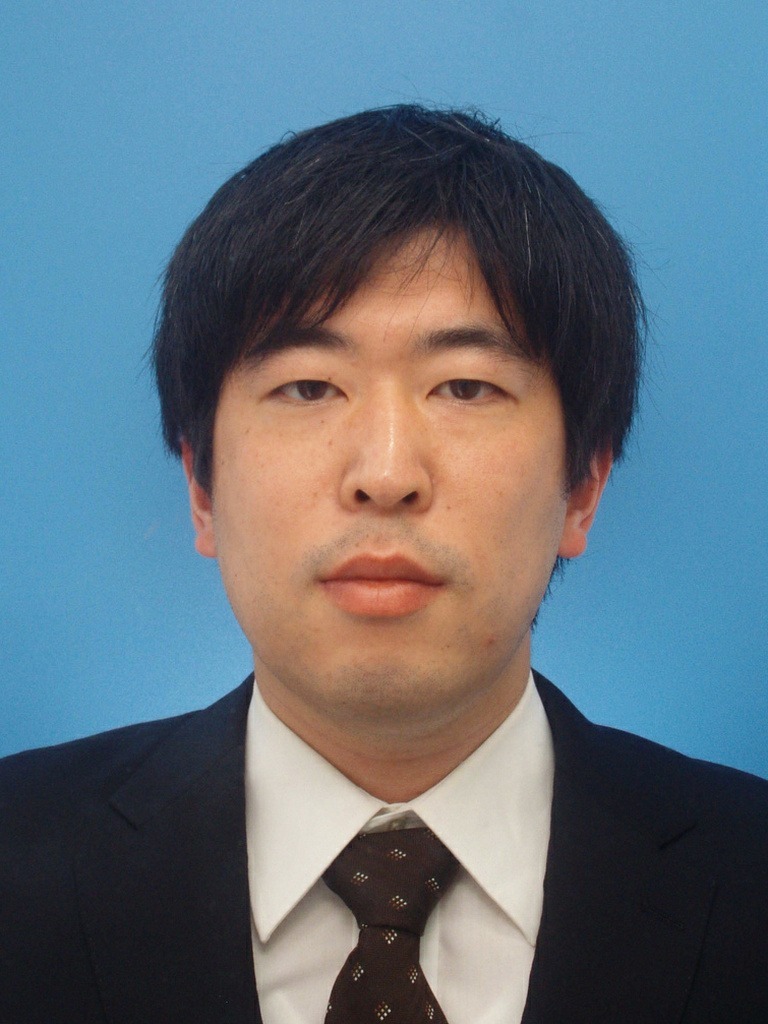}}]{\noindent Kenta Ito} (Graduate Student Member, IEEE) received the B.E. degree in electrical engineering from Toyama University, Japan, in 2019, and the M.E. degree in communication engineering from Osaka University, Japan, in 2021. He is currently pursuing the Ph.D. degree at the Graduate School of Engineering, Osaka University. His research interests include belief propagation, compressed sensing, signal processing, and wireless communications.
\end{IEEEbiography}

%






\vfill

\end{document}